\documentclass[a4paper,11pt]{article}
\usepackage{jheppub} % for details on the use of the package, please see the JINST-author-manual
\usepackage{lineno}
\usepackage{physics}
\usepackage{amssymb}
\usepackage{amsmath}
\usepackage{mathtools}
\usepackage{tikz}
\usetikzlibrary{chains}
\usepackage{mathtools}

\DeclarePairedDelimiter\floor{\lfloor}{\rfloor}

\tikzset{node distance=2em, ch/.style={circle,draw,on chain,inner sep=2pt},chj/.style={ch,join},every path/.style={shorten >=4pt,shorten <=4pt},line width=1pt,baseline=-1ex}

\author{Chao Ju}
\affiliation{Berkeley Center for Theoretical Physics and Department of Physics\\
University of California, Berkeley, CA 94720, U.S.A.}

\emailAdd{cju19@berkeley.edu}

\abstract{Classically, the ground states of $\mathcal{N}=4$ supersymmetric Yang-Mills theory on $\mathbb{R}\times S^3/\Gamma$ where $\Gamma$ is a discrete ADE subgroup of $SU(2)$ are represented by flat Wilson lines winding around the ADE singularity. By a duality relating such ground states to WZW conformal blocks, the ground state degeneracy cannot be lifted by quantum corrections. Using the superconformal index, we compute the supersymmetric Casimir energy of each flat Wilson line for $SU(2)$ SYM on different ADE singularities and find that the flat Wilson lines all have the same supersymmetric Casimir energy. We argue that this exact degeneracy is peculiar to $\mathcal{N}=4$ supersymmetry and show that the degeneracy is lifted when the number of supersymmetry is reduced. In particular, we uncover a surprising result for the ground state structure of the conformal $\mathcal{N}=2$ $SU(2)$ four-flavor theory on $S^3/\Gamma$. For $\mathcal{N}=4$ SYM, S-duality maps the ground state Wilson lines to ground state t' Hooft lines taking values in the Langlands dual group. We show that the supersymmetric Casimir energy of the t' Hooft line ground states is the same as the Wilson line ground states. This can be viewed as a ground state test of S-duality.}

\title{Exact Degeneracy of Casimir Energy for $\mathcal{N}=4$ Supersymmetric Yang-Mills Theory on ADE Singularities and S-Duality}

\begin{document}
\maketitle
\flushbottom

\section{Introduction}
\label{sec:intro}
Supersymmetric Yang-Mills theory in four dimension has a varying degree of applications depending on the number of supercharges the theory has. $\mathcal{N}=1$ SYM with flavor degrees of freedom (SQCD) is not only useful for phenomenological purposes~\cite{weinberg,dine,affleck,wittensymbreak}; it also provides a fertile ground for concrete examples of dualities~\cite{seiberg}. $\mathcal{N}=2$ SYM provided the first realization of quark confinement via monopole condensation~\cite{sw1}. A whole class of $\mathcal{N}=2$ theories can be engineered by putting M5 branes on Riemann surfaces, and the dualities bewteen the theories can be analyzed using the geometry of the Riemann surface~\cite{gaiotto}. For a review, see~\cite{tachikawa}. $\mathcal{N}=4$ SYM is a UV finite theory~\cite{weinberg}, and provides an explicit realization of holographic duality via the AdS/CFT correspondence~\cite{malda}. It is this UV finite theory that will concern us in this paper. 

Recent years saw a proliferation of index technology in understanding the strongly-coupled regime of supersymmetric Yang-Mills theories and dualities~\cite{romelsberger,kinney,aharony}, although the use of index dates back to~\cite{wittenindex}. For a review, see~\cite{rastelli} and~\cite{gadde} and the references therein. Using supersymmetric index, one can obtain more information on the theory if the underlying geometry has nontrivial first fundamental group and preserves some number of supersymmetry, since nontrivial flat Wilson lines from the vector multiplet (or from background flavor gauge fields) can wind around such geometry and divides the theory into different holonomy sectors. A salient example is the lens space supersymmetric index~\cite{razamat,benini,razamat2}. The particular lens space considered in these works is $L(r,1)$, or $S^3/\mathbb{Z}_r$, defined by identifying the point $(z_1,z_2)$ on $S^3$ with the point $(wz_1,w^{-1}z_2)$ where $w$ is the $r$th root of unity. There are two notable properties of this identification. First, the identification acts only on the holomorphic part of the coordinates, suggesting that one can choose some supersymmetry to be preserved. Second, the identification forms an abelian discrete subgroup of $SU(2)$ and acts separately on $z_1$ and $z_2$. There exist other discrete subgroups of $SU(2)$ that satisfy the first property but break the second (i.e. they are non-abelian). In fact, all discrete subgroups of $SU(2)$ have been classified and are found to correspond to simply-laced Dynkin diagrams~\cite{mckay}. The A-series corresponds to the aelian $\mathbb{Z}_k$ subgroups, and produces the lens space $L(k,1)$. The D-series corresponds to the non-abelian $\text{Dic}_k$ (binary dihedral) subgroups. The E-series corresponds to the double-cover of symmetry groups for the tetrahedron, the octahedron, and the icosahedron (respectively for E6, E7, and E8). Although modding out $S^3$ by the D and the E subgroups leads to fundamental groups that are non-abelian, it is nevertheless sensible to consider the supersymmetric index on these more complicated geometries because they can preserve the same number of supercharge as the lens space. In this paper, we will compute the supersymmetric index on these nontrivial geometries.

The motivation for computing the supersymmetric index on ADE singularities is as follows. In an earlier work~\cite{ganorandju}, we found a duality relating the ground state Hilbert space of $U(n)$ $N=4$ supersymmetric Yang-Mills theory quantized on $S^3/\Gamma$ to the Hilbert space of level $n$ Chern-Simons theory with gauge group $G(\Gamma)$ quantized on $T^2$. Here, $\Gamma$ is one of the ADE subgroups of $SU(2)$ and $G(\Gamma)$ is the corresponding gauge group given by the McKay correspondence~\cite{mckay}. Classically, the ground states of the SYM theory are classified by the homomorphism $\pi_1(S^3/\Gamma)\to SU(2)$ up to identification by the Weyl group and $U(n)$ conjugation. Because $S^3$ is simply connected, the classical ground states are simply Wilson lines taking values in $U(n)$ and satisfying the multiplication rule of the group $\Gamma$. The rest of the fields in the vector multiplet are all set to 0 because of the coupling to curvature~\cite{wittenads}. For a similar construction of this duality using a twisted SYM theory and with a focus on moduli space of instantons, see~\cite{vafawitten,nakajima}. For a slightly less general setting than ours, see ~\cite{dijkgraaf}.

In~\cite{ganorandju,ju}, detailed calculations show that the dimension of the ground state Hilbert space of the SYM theory (i.e. the number of classically flat Wilson lines) matches that of the corresponding Chern-Simons theory. On the Chern-Simons theory side, we know that the ground state degeneracy is a one-loop exact result~\cite{wittenjones} and that quantum corrections do not generate any potentials. We therefore expect that the same thing happens on the dual SYM side: the classically flat Wilson lines stay flat even quantum mechanically. There is no way to distinguish between them. 

This does not in general happen when supersymmetry is not present. Although in this paper we deal with flat Wilson lines taking discrete values according to the ADE group $\Gamma$, we give some examples using Wilson lines taking continuous values to support this claim. Without supersymmetry, toroidal compactification in general will dynamically generate a potential for the dilaton field and localizes it~\cite{polchinski}, the dilaton being analogous to the classically flat Wilson line. Another example where quantum corrections lift the classical degeneracy in flat Wilson lines is 2D Yang-Mills on a circle. The quantum mechanical partition function of this theory is~\cite{tachikawa2}
\begin{equation}
    Z=\sum_R e^{-TLc_2(R)}
\end{equation}
where $T$ and $L$ are the lengths of the time and the spatial circle, the sum is over the irreducible representations, and $c_2(R)$ is the quadratic Casimir of irreducible representation $R$. From the expression, one sees that classically flat Wilson lines now gain different amount of energy depending on the representation $R$.

Even with the help of supersymmetry, Wilson lines (dilatons) exhibit different behaviors depending on the amount of supersymmetry. Let us start with the example of $n$ coincident D4 branes in type IIA string theory with worldvolume in the 01234 directions. This system breaks half of the 32 supersymmetries and leads to the 5D $U(n)$ supersymmetry Yang-Mills theory on the D4 brane world volume. Now, we compactify the 4th directions of the D4 branes. We can adjust the value of the Wilson line by tuning the fourth component of the gauge field $A_4$. Doing so does not cost any energy classically, and we can conjugate $A_4$ to take value in the Cartan subalgebra of $\mathfrak{u}(n)$. That this does not cost any energy quantum mechanically can be seen by going to the T-dual picture, where the system now contains $n$ D3 branes separated along the 4th direction given by the value of $A_4$. The distance $X^4$ is given by the T-dual relation $X^4=2\pi \alpha'A_4$ where $\alpha'$ is the string length~\cite{polchinski}. This system also breaks half of the 32 supersymmetries, and no potential is generated for the classically flat Wilson line~\cite{polchinski}. In contrast, consider the system of a D4 brane with world volume in the 01234 directions and a D2 brane in the 034 directions in type IIA string theory. We further suppose that there is no separation between the D-branes in the 56789 directions. This system breaks one quarter of the 32 supersymmetries as opposed to one half, and it leads to $\mathcal{N}=2$ supersymmetry as measured in 4D. In one picture, the D2 brane dissolves in the D4 brane, leaving an RR flux on the D4 brane~\cite{polchinski}. In another picture, we compactify the 4th direction, and tune the Wilson line as before such that it is classically flat. In the T-dual picture, this separates the D3 brane and the D1 brane in the 4 direction, and it is further U-dual to the F1-D1 system. It is known that the F1-D1 system is unstable and forms a bound state~\cite{wittenpstring}. There is therefore an attractive force between the D3 brane and the D1 brane, causing the dilaton to gain a potential.

The lesson in the previous paragraph is that the number of supersymmetry can affect whether or not classically flat Wilson lines can gain a potential. In this work, we deal with discrete flat Wilson lines, and a priori there is no reason to expect such statement to hold for discrete flat Wilson lines. Nevertheless, as we will see in section~\ref{sec:susy}, the discrete flat Wilson lines can become nonflat when the number of supersymmetry is reduced from 4 to 2.

We define the one-loop correction to the $j$th ground state energy as the supersymmetric Casimir energy $E_j$ mentioned in~\cite{benini}. It is the nonsingular part of
\begin{equation}
\label{eq:casimirdef}
E_j = -\frac{1}{2} \lim_{\beta\to 0} \frac{\partial}{\partial\beta} \hat{I}_j
\end{equation}
where $\hat{I}_j$ is the single letter supersymmetric index of the theory for the $j$th ground state and $\beta$ is some fugacity parameter coupled to some suitable Hamiltonian (to be discussed later) that commutes with the supersymmetry subalgebra used in computing the index. In this work, we will show that for $\mathcal{N}=4$ SUSY
\begin{equation}
    E_1 = E_2 = ... = E_{q}
\end{equation}
where $q$ is the dimension of the classical ground state Hilbert space and $1$ corresponds to the trivial Wilson line (i.e. the identity). We show this for $N=4$ SYM with $SU(2)$ gauge group on different ADE singularities $\Gamma$, and conjecture that this relation holds for all $SU(N)$ or $U(N)$ gauge groups. This result can be viewed as a one-loop test of the duality proposed in~\cite{ganorandju,vafawitten,dijkgraaf}: The ground state Hilbert space of $N=4$ $SU(N)$ SYM on $S^3/\Gamma$ is dual to certain subspace of $G(\Gamma)$, level $N$ Chern-Simons theory on $T^2$. The reason we focus on the $SU(2)$ gauge group in this work is twofold. First, it is the simplest non-abelian gauge group and it makes the index computation tractable. Second, we want to compare our result with the conformal $\mathcal{N}=2$ four-flavor theory which also has gauge group $SU(2)$. A novelty of this work is that the single-letter supersymmetric index on the D-singularity is computed for the first time, complementing the A-singularity (lens space) result. 

One reason we choose to measure the supersymmetric Casimir energy using a Hamiltonian that commutes with the supersymmetry subalgebra is that we want to have an energy measure we can trust at strong coupling. At strong coupling, we can use S-duality to go to a weakly coupled theory and compute the supersymmetric Casimir energy of the flat t' Hooft lines. There will be two nontrivial tests of S-duality. First, the number of flat t' Hooft lines must match the number of the number of flat Wilson lines. Second, let us denote the supersymmetric Casimir energy of the $i$th flat t' Hooft lines by $\tilde{E}_i$. S-duality predicts
\begin{equation}
    \tilde{E}_1= \tilde{E}_2 =...=\tilde{E}_q = E_1=E_2=...=E_q
\end{equation} 
namely the flat t' Hooft lines must also be degenerate in supersymmetric Casimir energy as the flat Wilson lines. These two facts will be checked in section~\ref{sec:sduality}.

This work differs from other works in the literature in terms of 1) the underlying geometry, 2) the number of supercharges, 3) the definition of the supersymmetric Casimir energy, and 4) the scope of the problem. Our definition of the supersymmetric Casimir energy follows that of~\cite{benini} and~\cite{razamat2} but not that of~\cite{assel}. In~\cite{assel}, the focus is on $\mathcal{N}=1$ theory on the nonsingular $S^3\times S^1$, and the supersymmetric Hamiltonian is chosen to have a twist by the $R$ supercharge because of the holonomy around the time circle. In terms of the scope of the problem, we do not discuss the full SUSY index; we only care about the Casimir part of it. In addition, We do not discuss the compactification of 4D theory to 3D, where for certain bad 4D $\mathcal{N}=2$ theories~\cite{gaiottowitten} the 3D SUSY index may diverge~\cite{giacomelli} due to certain monopole operators. Note that~\cite{giacomelli} considers non-singular geometry only. For compacitifcation of the $\mathcal{N}=2$ and $\mathcal{N}=1$ index from 4D to 3D on a singular geometry (e.g. lens space), see~\cite{benini}.

The paper is organized as follows. In section~\ref{sec:susy}, we analyze the $\mathcal{N}=4$ superconformal algebra and set the convention we use in defining the supersymmetric index. In section~\ref{sec:a} and section~\ref{sec:d}, we compute the supersymmetric Casimir energy for $\mathcal{N}=4$ $SU(2)$ SYM on A- and D-singularity, respectively. We set up the calculation for the E-singularity in section~\ref{sec:e}. In section~\ref{sec:comp1}, we compare our result to theories with less supersymmetry, namely $\mathcal{N}=2$ supersymmetry. We find that generically, only $\mathcal{N}=4$ theory has an exact degeneracy of supersymmetric Casimir energy. This makes sense from the holographic duality perspective, since the duality proposed in~\cite{ganorandju,vafawitten,dijkgraaf} has a D3(M5) brane realization where the number of supercharges is 16 (see section 9 of~\cite{ju} for a derivation). No such duality exists for supersymmetry less than $\mathcal{N}=4$ except for some particular $\mathcal{N}=2$ theories (class-S). In particular, we will find a surprising result that there is no ground state degeneracy for the conformal $\mathcal{N}=2$ four-flavor theory on $S^3/\Gamma$: not all classically flat Wilson lines are created equal for this theory. Finally, in section~\ref{sec:sduality}, we compute the supersymmetry Casimir energy for the t' Hooft lines to give yet another test of S-duality.

\section{Analysis of Supersymmetry}
\label{sec:susy}
As discussed earlier, the approach we take to compute the supersymmetric Casimir energy is through the supersymmetric index calculation. Since the index calculation hinges on understanding the $N=4$ superconformal algebra, we take a moment to review the algebra in this section. We shall mostly follow the convention of~\cite{kinney}.

Since we are dealing with $\mathcal{N}=4$ supersymmetry, the superalgebra of interest has an $SU(4)$ R-symmetry. Without the conformal part of the algebra, the superalgebra is closed under the fermionic symmetry generators $Q^{\alpha i}$, $\bar{Q}^{\dot{\alpha}}_i$, the Lorentz symmetry generator $J^{\alpha}_{2\beta}$, $J^{\dot{\alpha}}_{2\beta}$, the $SU(4)$ R-symmetry generator $R^i_{j}$, and the translation generator $P^{\alpha\dot{\beta}}$. Here, $\alpha, \dot{\alpha}$ are $SU(2)_L$ and $SU(2)_R$ indices coming from the Lorentz group, and $i=\{1,2,3,4\}$ is the $SU(4)$ R-symmetry index. Because the theory is conformal even at the quantum level, one can enlarge the superalgebra by adding in the conformal algebra. The enhancement produces the fermionic counterpart of $Q$ and $\bar{Q}$: $S_{\alpha i}$, $\bar{S}^i_{\dot{\alpha}}$, the bosonic counterpart of $P$: $K^{\alpha\dot{\beta}}$, and the dilitation operator $D$. These form the superalgebra $SU(2,2|4)$. 

In radial quantization, the S generators and Q generators are Hermitian conjugate of each other, so that we have the positive definite anticommutator~\cite{kinney}
\begin{equation}
\label{eq:susyalgebra}
    \{Q^\dagger_{\alpha i}, Q^{\beta j}\}=  \delta^j_i J^\beta_{1\alpha} +\delta^\beta_\delta R^j_i + \delta^j_i\delta^\beta_\alpha \frac{D}{2}
\end{equation}
A similar relation holds if we replace $Q$ by $\bar Q$, the undotted indices by dotted indices, and $J_1$ by $J_2$. A comprehensive list of other (anti)commutators can be found in the appendix of~\cite{kinney}. The above anticommutator will be enough for our purposes. 

To define the index, we need to pick out a particular $Q, Q^\dagger$ pair and look for states annihilated by $\{Q^\dagger,Q\}$. Let us pick $Q=Q^{-1/2,1}$, which according to equation~\eqref{eq:susyalgebra} gives the anticommutator 
\begin{equation}
\label{eq:delta}
    2\{Q^\dagger,Q\} = D- 2J_1 - \left(\frac{3}{2}R_1 + R_2 + \frac{1}{2}R_3\right)
\end{equation}
where $R_1, R_2, R_3$ are the maximally commuting $SU(4)$ charges and $J_1$ is the $SU(2)_L$ charge. States that are annihilated by the above anticommutator form short BPS multiplets whose contribution to the index
\begin{equation}
\label{eq:bareindex}
    Z=\tr(-1)^F
\end{equation}
does not change as the coupling constant is varied, providing a reliable probe of the strongly-coupled regime of the theory~\cite{wittenindex}\cite{kinney}. Nevertheless, equation~\eqref{eq:bareindex} is not the most general quantity that is protected when one restricts the partition function in the Hilbert space annihilated by the anticommutator~\eqref{eq:delta}. One can also add in other quantum numbers that commute with the superalgebra generated by $Q$ and $Q^\dagger$. The commuting subalgebra is $SU(2,1|3)$, whose bosonic Cartan elements are
\begin{equation}
    D+J_1, J_2, R_2, R_3
\end{equation}

Therefore, if one wants to obtain maximal information on the protected spectrum of the theory, one can compute
\begin{equation}
\label{eq:indexwithfugacities}
    Z=\tr((-1)^F e^{-\beta(D+J_1 + \Omega_2 J_2 + tR_2 +u R_3)})
\end{equation}
where $\beta$, $\beta\Omega_2$, $\beta t$, $\beta J$ are the bookkeeping parameters (fugacities) that help us distinguish states with different quantum numbers. They also help regulate\footnote{Since the $N=4$ theory is a UV finite theory, we will see that there will be no UV divergence in the index calculation.} the infinite sum in~\eqref{eq:bareindex}. Since we are ultimately interested in the supersymmetric Casimir energy, we set $\Omega_2, t, u$ to be zero and focus only on the $D+J_1$ part. Another reason for setting $\Omega_2$ to zero is that nonabelian singularities such as the D- and the E-singularity do not commute with $J_2$, so $J_2$ is a bad quantum number\footnote{As an example, the defining two-dimensional representation of the reflection element in the $\text{Dic}_k$ group is 
\[
\begin{pmatrix}
    0 & 1 \\
    -1 & 0
\end{pmatrix}
\]
whose action turns $\partial_{++}^{n_1}\partial_{+-}^{n_2}$ into $(-1)^{n_2}\partial_{++}^{n_2}\partial_{+-}^{n_1}$, thus changing the $J_2$ value of this operator from $n_1-n_2$ to $n_2-n_1$. This situation does not occur for the A-singularity (lens space) which is abelian.
}.

For this work, it is sensible to define the ``supersymmetric energy'' as $D+J_1$ for four reasons. First, $D$ generates time translation in the radially quantized picture, so to define a Hamiltonian we need a quantity that contains $D$. Second, $D+J_1$ commutes with $\{Q^\dagger, Q\}$, so an index calculation involving this quantity gives a well-defined answer, independent of the value of the coupling constant. This property is crucial for our S-duality calculation in section~\ref{sec:sduality}, since we need a quantity that can be meaningfully compared in two different strong-weak duality frames. Had we chosen a Hamiltonian that does not commute with $\{Q^\dagger, Q\}$, we would not have been able to have a good test of S-duality. Third, as we will check later, $D+J_1$ is positive definite on the BPS states annihilated by $\{Q^\dagger,Q\}$. Finally, $D+J_1$ is what matters when we take the so-called MacDonald limit~\cite{macdonald} of the index, setting $t=0$. In this limit, we are counting states that are annihilated by one more pair of supersymmetry charges. If the theory contains states that are annihilated by at least two pairs of supersymmetry charges, the index must have a well-defined MacDonald limit $t=0$.

Having discussed why we chose to define the supersymmetric energy as $D+J_1$, we now analyze how to compute the Casimir part of it. Because the index is independent of the coupling constant, the computation of 
\begin{equation}
    \label{eq:index1}
    Z=\tr((-1)^F e^{-\beta(D+J_1)})
\end{equation}
can be carried out by doing a twisted path integral on the free theory, the result being some determinant factors~\cite{benini}. The result is trivially one-loop exact (since the coupling constant can be safely set to 0 without affecting the result), and the supersymmetric Casimir energy corresponding to $D+J_1$ is given by~\cite{benini} as in equation~\eqref{eq:casimirdef}. A quick way to see why~\eqref{eq:casimirdef} yields the supersymmetric Casimir energy without going through the path integral derivation is to recall that this is how the $-1/24$ normal ordering constant arises in bosonic string theory~\cite{polchinski}. There, the single letter ``index'' is simply a sum over $e^{-\beta L_0}$ on the vacuum and its Virasoro descendent where $L_0$ is the Cartan element of the Virasoro generators.

There is one more ingredient we need to discuss before computing the index. In our setting, the ground states are labeled by different flat Wilson lines $g_j$ taking values in $SU(2)$ and furnishing a representation of $\pi_1(S^3/\Gamma)=\Gamma$. To count the contribution to the single letter index, we gauge away the flat Wilson lines at the cost of introducing a twist to other fields when transported around a nontrivial loop. To see this, let the nontrivial loop starting from $a$ and ending at $b$ be $l$ and let the gauge field $A$ in the $j$th flat Wilson line sector be denoted by $A^{(j)}$. Note that point $a$ and point $b$ are identified under some group element $\gamma\in \Gamma$: $b=\gamma a$. The Wilson line wrapping $l$ is
\begin{equation}
    g_j(\gamma) = P\exp i\oint_{l}  A^{(j)} \quad \in SU(2)
\end{equation}

The notation $g_j(\gamma)$ is suggestive, as the Wilson lines are representations of the group $\Gamma$. Here, $g_j$ is the $j$th representation of the specific group element $\gamma\in\Gamma$. This Wilson line can be gauged away by using a non-periodic gauge transformation $U\in SU(2)$ such that
\begin{align*}
    U(a) &= 1 \\
    U(\gamma a) &= g_j(\gamma)
\end{align*}

The effect of this nonperiodic gauge transformation on a field $\phi$ that transforms in the fundamental of $SU(2)$ is such that, when $\phi$ is transported from $a$ to $\gamma a$, it is multiplied by $U(\gamma a)=g_j(\gamma)$. For a field $\Phi$ that transforms in the adjoint of $SU(2)$, the field becomes $g_j(\gamma)^{–1} \Phi g_j$ as it is transported around the loop. In the $N=4$ SUSY theory, we only have adjoint fields, so we write down the transformation rule for a general adjoint field $\Phi$ around a loop $l$ with flat Wilson line $g_j(\gamma)$:
\begin{equation}
    \Phi(\gamma a) = g_j(\gamma)^{-1}\Phi(a) g_j(\gamma)
\end{equation}

The group $\Gamma$ acts geometrically on the underlying space. This geometric action induces an action on the fields $\Phi$ such that the field values at $a$ and $\gamma a$ are related. Define this induced action as
\begin{equation}
    \Phi(\gamma a) = \gamma \Phi(a)
\end{equation}

Using this, we see from the previous equation that the adjoint field $\Phi$ must satisfy the constraint
\begin{equation}
    \label{eq:constraint}
    \gamma \Phi = g_j^{-1}(\gamma)\Phi g_j(\gamma)
\end{equation}

In the Wilson line sector $j$, for a field $\Phi$ to enter into the single letter index counting, it must satisfy the above constraint for all $\gamma\in \Gamma$. If the above equation holds for all the $k$ generators $\gamma_1,...\gamma_k$ of $\Gamma$, it holds for all $\gamma\in \Gamma$. Therefore, we effectively have only $k$ constraints for each $j$. As we will see, the $k$ value for the $A$, $D$, $E$ groups are 1, 2, and 2, respectively. 

The strategy for computing the single letter index is now clear. By the state-operator correspondence, We start with a field $\Phi$ that is annihilated by $\{Q^\dagger,Q\}$ and act on it using the derivatives $\partial_{+\pm}$ which are also annihilated by $\{Q^\dagger, Q\}$, keeping the BPS condition. We project out all the operators that do not satisfy the constraint~\eqref{eq:constraint}. For those that do, we add to the single letter index their Boltzmann weight
\begin{equation}
\label{eq:weight}
    (-1)^F e^{-\beta(D+J_1)} = (-1)^F t^{2(D+J_1)}
\end{equation}
in which, following the convention of~\cite{benini}, we define
\begin{equation}
    t= e^{-\beta/2}
\end{equation}

It is useful to have a list of fields that satisfy the BPS condition so that we could view their quantum numbers $D, j_1, j_2$. Such a list of fields and their quantum numbers is provided by~\cite{kinney}. Although the $j_2$ quantum number does not explicitly appear in the constraint~\eqref{eq:constraint}, it enters implicitly on the left hand side (the geometric part of the action). The reason is that we chose to embed $\Gamma$ in $SU(2)_R$ which breaks half of the supersymmetry. Therefore, $\Gamma$ can act nontrivially on fields with a nonzero $j_2$ value. We will write down the action explicitly for each of the ADE groups we encounter later.

\begin{table}[htbp]
\centering
\begin{tabular}{|c|c|}
    \hline
    Letter & $(-1)^F [E, j_1, j_2]$  \\
    \hline
    X, Y, Z & $[1,0,0]$ \\
    \hline
    $\psi_{+,0;-++},\psi_{+,0;+-+},\psi_{+,0;++-}$ & $-[3/2,1/2,0]$ \\
    \hline
    $F_{++}$ & $[2,1,0]$ \\
    \hline
    $\partial_{++}\psi_{0,-;+++} +\partial_{+-}\psi_{0,+;+++}=0$ & $[5/2,1/2,0]$ \\
    \hline
    $\psi_{0,\pm;+++}$ & $-[3/2,0,\pm 1/2]$ \\
    \hline
    $\partial_{+\pm}$ & $[1,1/2,\pm 1/2]$ \\
    \hline
\end{tabular}
\caption{A list of operators that satisfy the BPS condition and their $E, j_1, j_2$ quantum numbers. The $R_1, R_2, R_3$ quantum numbers are omitted and can be found in~\cite{kinney}. \label{table:n=4bpstable}}
\end{table}

The operators listed in this table are all BPS. In the fourth operator, the minus sign for fermion is canceled out by the Dirac equation since we want to subtract its contribution from the index to avoid overcounting. With the exception of the last two operators, everything else has zero $j_2$ quantum number. We are now ready to compute the single letter index and the supersymmetric Casimir energy on each ADE geometry. 

\section{A-Singularity}
\label{sec:a}
The index calculation for the A-singularity (lens space) has been done before~\cite{razamat,benini,razamat2} for $\mathcal{N}=2$ theories. We here redo the calculation for the $\mathcal{N}=4$ theory in a way that can be generalized to the D- and the E-singularity. In this section, we will discuss the logic of our computation and relegate the details to the appendix. 

The A-series are the groups $\Gamma=\mathbb{Z}_k$ whose group presentation is
\begin{equation}
    r^k=e
\end{equation}

We first discuss the geometric action and then the Wilson line action. The geometric action, as discussed in the previous section, has to do with the generators of the group. For $\mathbb{Z}_k$ there is a single generator which we call $r$. According to the McKay correspondence~\cite{mckay}, the defining geometric action of $r$ on $SU(2)$ doublet is~\cite{aspinwall}
\begin{equation}
    r=\begin{pmatrix}
        w & 0 \\
        0 & w^{-1}
    \end{pmatrix}
\end{equation}
where $w = \exp(2\pi i/k)$. We would now like to understand how $r$ acts geometrically on various BPS operators. From the Table~\ref{table:n=4bpstable}, we see that $\partial_{+\pm}$ transforms as a doublet under $SU(2)$, since their $j_2$ quantum numbers are $\pm 1/2$. Thus, under $r$, they transform as
\begin{equation}
    \begin{pmatrix}
        \partial_{++} \\
        \partial_{+-} 
    \end{pmatrix}
    \to 
    \begin{pmatrix}
        w & 0 \\
        0 & w^{-1}
    \end{pmatrix}
    \begin{pmatrix}
        \partial_{++} \\
        \partial_{+-} 
    \end{pmatrix}
\end{equation}

This suggests that, if we start with a field $\Phi$ that has $j_2=0$ and acts on it by $\partial_{++}^{n_1}\partial_{+-}^{n_2}$, it will transform geometrically as
\begin{equation}
    \label{eq:ar1}
    r(\partial_{++}^{n_1}\partial_{+-}^{n_2}\Phi) = w^{n_1-n_2} \partial_{++}^{n_1}\partial_{+-}^{n_2}\Phi
\end{equation}

The only fields in the theory that has $j_2\neq 0$ are $\psi_{0,\pm;+++}$, which have $j_2=\pm 1/2$. This suggests that, just as $\partial_{+\pm}$, $\psi_{0,\pm;+++}$ also transform as a doublet under $r$. Denote such a doublet field by $\Phi_\mu$, where $\mu=\pm 1$. We have
\begin{equation}
    \label{eq:ar1}
    r(\partial_{++}^{n_1}\partial_{+-}^{n_2}\Phi_\mu) = w^{n_1-n_2+\mu} \partial_{++}^{n_1}\partial_{+-}^{n_2}\Phi_\mu
\end{equation}

This concludes the discussion of the geometric action of $r$ on various operators. We now discuss the action of the Wilson lines which take value in $SU(2)$ representations of $\mathbb{Z}_k$. There are $k$ irreducible representations for the group $\mathbb{Z}_k$. From the $k$ irreducible representations we can build up $\floor*{k/2}$ $SU(2)$ representations of the form
\begin{equation}
    g_n(r) = 
    \begin{pmatrix}
    w^n & 0 \\
    0 & w^{-n}
    \end{pmatrix}
    \quad j=0,1,...,\floor*{k/2}
\end{equation}

A truncation at $n=\floor*{k/2}$ happens because of the Weyl group symmetry, which exchanges the diagonal elements of the $SU(2)$ matrix. The $n$th Wilson line $g_n$ acts on the $SU(2)$ adjoint fields by conjugation. Denote a general adjoint field by $\Phi_{l}$, $l=-1,0,1$, so that $\Phi_0 \in \mathbb{R}$ and $\Phi_{1}=\Phi^{*}_{-1} 
\in \mathbb{C}$. Since the Wilson line does not act on the geometric part of the field, we suppress the $j_2$ doublet index $\mu$ if there is any. Assembling the three components of the adjoint fields into a matrix:
\begin{equation}
\Phi=
    \begin{pmatrix}
    \Phi_0 & \Phi_1 \\
    \Phi_{-1} & -\Phi_0
    \end{pmatrix}
\end{equation}
We see that under $\Phi \to g_n(r) \Phi g_n^{-1}(r)$ the components transform as
\begin{align*}
    g_n(r)(\Phi_0, \Phi_1, \Phi_{-1})g^{-1}_n(r) = (\Phi_0, w^{2n} \Phi_1, w^{-2n}\Phi_{-1}) 
\end{align*}
as is familiar from the theory of angular momentum.

Consider the $n$th ground state Wilson line. We would like to compute the supersymmetric Casimir energy from the single letter index, which, by the state operator correspondence, has contribution from BPS derivatives $\partial_{++}^{n_1}\partial_{+-}^{n_2}$ on BPS fields. Without the $\mathbb{Z}_k$ action, all nonnegative $n_1$ and $n_2$ lead to valid single letter operator that can contribute to the index. The $\mathbb{Z}_k$ Wilson line will, however, project out many states. The goal is to figure out the constraint on $n_1$ and $n_2$. We first discuss the case where the BPS field $\Phi_l$ has $j_2=0$. Here, $l=-1,0,1$ is the $SU(2)$ adjoint index. The most general linear combination of single letter operators is
\begin{equation}
    \Psi= \sum_{n_1n_2l} C_{n_1n_2l}\partial_{++}^{n_1}\partial_{+-}^{n_2} \Phi_l
\end{equation}
where $n_1,n_2\geq 0$. We would like to understand for what values of $n_1,n_2,l$ is $C_{n_1n_2l}$ nonvanishing, because that would indicate a contribution to the single letter index. Applying the constraint equation~\eqref{eq:constraint}, we need to have
\begin{equation}
    r\Psi = g_n(r) \Psi g_n(r)^{-1}
\end{equation}
where the left hand side is the geometric action as in equation~\eqref{eq:ar1}. Substituting in the expression for $\Psi$, we can turn the above equation into
\begin{equation}
    \sum_{n_1n_2l} w^{n_1-n_2}C_{n_1n_2l}\partial_{++}^{n_1}\partial_{+-}^{n_2} \Phi_l =\sum_{n_1n_2l} w^{2nl} C_{n_1n_2l} \partial_{++}^{n_1}\partial_{+-}^{n_2} \Phi_l
\end{equation}

Assuming the states corresponding to the operators are orthogonal as in 2D CFT, we obtain
\begin{equation}
    \exp(\frac{2\pi i (n_1-n_2)}{k}) C_{n_1n_2l} = \exp(\frac{2\pi inl}{k}) C_{n_1n_2l}
\end{equation}
which suggests that $C_{mnl}$ is nonvanishing only when
\begin{equation}
    n_1-n_2-2nl = 0 \mod{k}
\end{equation}

There are three cases to consider $l=0$, $l=1$, and $l=-1$, although the last two should lead to the same contribution to the index by symmetry. When $l=0$, we need to sum over all $n_1,n_2\geq 0$ such that 
\begin{equation}
\label{eq:nmconstraint}
    n_1-n_2=0\mod{k}
\end{equation}
The contribution of $\partial^{n_1}_{++}\partial^{n_2}_{+-}$ to the Boltzmann weight~\eqref{eq:weight} is $t^{3n_1+3n_2}$, so we need to compute the sum
\begin{equation}
    F^{n}_{\mathbb{Z}_k,j_2=0}\equiv \sum_{n_1n_2} t^{3n_1+3n_2}
\end{equation}
in which $n_1$ and $n_2$ obey the constraint~\eqref{eq:nmconstraint}. Assume that $k$ is an even number, it is easy to see that the constrained sum is
\begin{align}
    F^{n}_{\mathbb{Z}_k,j_2=0} &= \frac{3(1+t^{3k})}{(1-t^6)(1-t^{3k})}, \quad n=0, k/2 \\
    F^{n}_{\mathbb{Z}_k,j_2=0} &= \frac{1+t^{3k}+2(t^{6n}+t^{3k-6n})}{(1-t^6)(1-t^{3k})}, \quad n\neq 0, k/2
\end{align}

In appendix~\ref{sec:appendix1}, the constrained sum $F'^n_{\mathbb{Z}_k,j_2=\pm 1/2}$ for $j_2=\pm 1/2$ fields are computed as in equation~\eqref{eq:appendixzkf1} and~\eqref{eq:appendixzkf2}. The final step is to add in the contribution of the BPS fields. According to Table~\ref{table:n=4bpstable}, the single letter index for the $n$th ground state Wilson line is thus
\begin{align}
    \label{eq:zkindex1}
    \hat{I}^n_{\mathbb{Z}_k} &= F^{n}_{\mathbb{Z}_k,j_2=0} (3t^2-3t^4+2t^6)-F^{n}_{\mathbb{Z}_k,j_2=\pm 1/2} t^3, \quad n=0,k/2 \\
    \label{eq:zkindex2}
    \hat{I}^n_{\mathbb{Z}_k} &= F'^{n}_{\mathbb{Z}_k,j_2=0} (3t^2-3t^4+2t^6)-F'^{n}_{\mathbb{Z}_k,j_2=\pm 1/2} t^3, \quad n\neq 0,k/2\\
\end{align}

A similar result can be obtained for the case $k$ is odd, as is done in appendix~\ref{sec:appendix1}. 

To obtain the supersymmetric Casimir energy, we can expand equations~\eqref{eq:zkindex1} and~\eqref{eq:zkindex2} to first order in $\beta$:
\begin{align}
    \hat{I}^n_{\mathbb{Z}_k} &=3-\frac{4\beta}{3k} + O(\beta^2),\quad n=0,k/2 \\
    \hat{I}^n_{\mathbb{Z}_k} &=1-\frac{4\beta}{3k} + O(\beta^2),\quad n\neq 0,k/2
\end{align}

There are two notable features. First, there are no terms divergent in $\beta$ as we take $\beta\to 0$. This is expected as the $\mathcal{N}=4$ theory is a UV finite theory. This feature does not occur when the number of supersymmetry is less than four, as we will see later. Second, applying the supersymmetric Casimir energy formula~\eqref{eq:casimirdef}, we find that the ground states are all degenerate:
\begin{equation}
\label{eq:wilsonsolutionzk}
    E^n_{\mathbb{Z}_k} = \frac{2}{3k} \quad n=0,1,...,k/2
\end{equation}

The exact value of the Casimir energy does not concern us, because in the path integral derivation~\cite{benini} there is an overall constant one is free to shift. The important point is that the Casimir energy is the same for all ground states. Just as there are no terms divergent as $\beta\to 0$, the ground state degeneracy is also a special feature for $\mathcal{N}=4$ supersymmetry and does not generically hold for less supersymmetric theories. Before we compare with theories with less supersymmetry, in the next section we shall compute the Casimir energy for the $\text{Dic}_k$ group where we have to confront with nonabelian geometric actions and nonabelian Wilson lines. We will see that the same degeneracy holds for $\text{Dic}_k$ group as well. This lends strong support to the Chern-Simons/SYM duality mentioned in section~\ref{sec:intro}.

\section{D-Singularity}
\label{sec:d}
The presentation for the $\text{Dic}_k$ (binary dihedral) group is
\begin{equation}
\label{eq:dicgroup}
    r^{2k}=e, \quad s^2=r^k, \quad s^{-1}rs=r^{-1}
\end{equation}
which is reminiscent of the dihedral group of the symmetry of a $2k$-gon. The difference here is that the reflection $s$ does not square to the identity. Instead, it squares to a central element of order two. The group $\text{Dic}_k$ contains $4k$ elements generated by $r$ and $s$. The geometric action of $r$ and $s$ on $SU(2)$ doublets is~\cite{aspinwall}
\begin{align*}
    r&= \begin{pmatrix}
        w & 0 \\
        0 & w^{-1}
    \end{pmatrix} \\
    s&=\begin{pmatrix}
        0 & 1 \\
        -1 & 0
    \end{pmatrix}
\end{align*}
where $\omega=\exp(\pi i/k)$. The novelty here compared with the lens space case is that there is a generator which acts geometrically in a nonabelian way.

The ground state Wilson lines are $SU(2)$ representations of $\text{Dic}_k$. We would like to know how many ground state Wilson lines there are and their explicit forms. The group representation of $\text{Dic}_k$ behaves differently depending on whether $k$ is even or odd. Here we focus on the case when $k\in 2\mathbb{N}$, since the odd case can be worked out similarly. In general, let the number of $SU(q)$ ground state Wilson lines for $\text{Dic}_k$ be $a_q$. The generating function $\Phi(z)$ for computing $a_q$ worked out in~\cite{ju} as
\begin{align}
\label{eq:generating}
    \Phi_{\text{Dic}_k}&\equiv 1+ a_1z^1 + a_2z^2+a_3z^3+... \\
    &= \frac{1}{4}\left(\frac{1}{(1-z)^4(1-z^2)^{k-1}} + \frac{2}{(1-z^2)^2(1-z^2)^{k/2}(1+z^2)^{k/2-1}} +\frac{1}{(1-z^2)^2(1-z^2)^{k-1}} \right) \nonumber
\end{align}

Since we are interested in the $SU(2)$ representation, we expand the generating function to second order in $z$ and find
\begin{equation}
    a_2 = 4+ \frac{k}{2}, \quad k\in 2\mathbb{N}
\end{equation}

We can confirm this formula by working out explicitly the $SU(2)$ representations. Since we only care about the representation of the two generators $r$ and $s$, we write out the $SU(2)$ representations for $r$ and $s$ only~\cite{ju}. First, we have four abelian $SU(2)$ representations:
\begin{align}
\label{eq:wilsondicksol1}
    g_1(r)&=\begin{pmatrix}
        1 & 0 \\
        0 & 1
    \end{pmatrix} \quad
    g_1(s)=\begin{pmatrix}
        1 & 0 \\
        0 & 1
    \end{pmatrix} \\
    \label{eq:wilsondicksol2}
    g_2(r)&=\begin{pmatrix}
        1 & 0 \\
        0 & 1
    \end{pmatrix} \quad
    g_2(s)=-\begin{pmatrix}
        1 & 0 \\
        0 & 1
    \end{pmatrix} \\
    \label{eq:wilsondicksol3}
    g_3(r)&=-\begin{pmatrix}
        1 & 0 \\
        0 & 1
    \end{pmatrix} \quad
    g_3(s)=\begin{pmatrix}
        1 & 0 \\
        0 & 1
    \end{pmatrix}\\
    \label{eq:wilsondicksol4}
    g_4(r)&=-\begin{pmatrix}
        1 & 0 \\
        0 & 1
    \end{pmatrix} \quad
    g_4(s)=-\begin{pmatrix}
        1 & 0 \\
        0 & 1
    \end{pmatrix}
\end{align}

In addition to the four abelian Wilson lines, we also have $k/2$ nonabelian Wilson lines (labled by a prime on $g$ in the following) given by
\begin{align}
\label{eq:wilsondicksol5}
g_n'(r)&=\begin{pmatrix}
\exp(n\pi i/k) & 0 \\
0 & \exp(-n\pi i/k) 
\end{pmatrix} \quad n=1,3,5,...,k-1 \\
g'_n(s)&=
\begin{pmatrix}
0 & 1 \\
-1 & 0
\end{pmatrix} \quad n=1,3,5,...,k-1 \nonumber
\end{align}

This shows that the number of $SU(2)$ representation of $\text{Dic}_k$ is indeed $4+k/2$ for $k$ even. 

We can work out the single letter index in the same way we did in the previous section for the lens space, except for each Wilson line we have two constraints given by the two generators instead of one. The single letter indices for the four abelian Wilson lines must be the same, because being $\pm 1$ they all act in the same way in the adjoint. We leave the details in the appendix, and quote the result here:
\begin{align*}
    \hat{I}^{\text{abelian}}_{\text{Dic}_k} &= 3\left(\frac{t^{6k}}{(1-t^6)(1-t^{6k})}+\frac{1}{1-t^{12}}\right) (3t^2-3t^4+2t^6) - 3 \frac{t^3+t^{6k-3}}{(1-t^6)(1-t^{6k})} t^3 \\
    &= 3 -\frac{\beta}{3k} + O(\beta^2)
\end{align*}
from which we can use~\eqref{eq:casimirdef} to read off the supersymmetric Casimir energy:
\begin{equation}
\label{eq:diccasimir}
    E^{\text{abelian}}_{\text{Dic}_k} = \frac{1}{6k}
\end{equation}

The $k/2$ nonabelian ground states have different single letter index labeled by $n$. From the appendix, they are given by
\begin{align*}
    \hat{I}^{n}_{\text{Dic}_k} &= \left(\frac{t^{6k}+t^{6n}+t^{6k-6n}}{(1-t^6)(1-t^{6k})}+\frac{t^6}{1-t^{12}}\right) (3t^2-3t^4+2t^6) \\
    &- \frac{t^{6n+3}+t^{6n-3}+t^{6k-3-6n}+t^{6k+3-6n}+t^3+t^{6k-3}}{(1-t^6)(1-t^{6k})} t^3 \\
    &= -\frac{\beta}{3k} + O(\beta^2)
\end{align*}

From the above expression, we see that the single letter index is different for each nonabelian Wilson line labeled by $n$, but the expansion in $\beta$ shows that they all agree to first order. This suggests that the $k/2$ nonabelian ground states all have the same Casimir energy
\begin{equation}
\label{eq:wilsonsolutiondick}
    E^{n}_{\text{Dic}_k} = \frac{1}{6k}, \quad n=1,3,5,...,k-1
\end{equation}

Comparing this with the supersymmetric Casimir energy of the abelian Wilson lines in equation~\eqref{eq:diccasimir}, we come to the conclusion that all $4+k/2$ ground states are degenerate, just like the case for lens space. In addition, the group $\text{Dic}_1$, being secretly abelian, is isomorphic\footnote{To see this, set $k=1$ in the group structure relations~\eqref{eq:dicgroup}.} to $\mathbb{Z}_4$, so putting $k=1$ in the above equation we should reproduce the $S^3/\mathbb{Z}_4$ result in equation~\eqref{eq:wilsonsolutionzk}. This is easily confirmed by setting $k=4$ in~\eqref{eq:wilsonsolutionzk}, which results in a Casimir energy of $1/6$. This provides a consistency check on our formalism.

\section{E-Singularity}
\label{sec:e}
In this section we comment on the computation of the supersymmetric Casimir energy for ground states on $S^3/E_k$ ($k=6,7,8$) where $E_k$ is any of the three symmetry groups for the platonic solids, also known as 2T (binary tetrahedral group), 2O (binary octahedral group), and 2I (binary icosahedral group). We will not be able to prove, like we did for the A- and the D-singularity, the exact degeneracy of supersymmetric Casimir energy for the E-singularity by a direct computation. Instead, we will start with a general discussion and use the specific example of $E_6$ to lay out the strategy for computing the single letter index on $S^3/E_6$. Although there is not any substantial result from this section, we feel that it is nevertheless beneficial to explicitly list the properties of a particular exceptional group ($E_6$) since the representation theory of $E_k$ may not be as familiar to many physicists. Another reason for writing this section is that the reader may use the information from here to come up with their own way to compute the single letter index on $S^3/E_k$.

Like $\text{Dic}_k$, all three $E_k$ groups have two generators. However, it is convenient to write down the group presentation for $E_k$ using three dependent generators $r,s,t$:
\begin{equation}
    r^2=s^3=t^{k-3}=rst = -1, \quad k=6,7,8
\end{equation}
where $-1$ denotes the central element of the group of order 2. It is easy to see that one of the generators can be expressed using the other two independent ones. 

In the following, we focus on the simplest of the three groups, $E_6$, the binary tetrahedral group of order 24. By the McKay correspondence~\cite{mckay}, there are three 1-dimension irreducible representations, three 2-dimensional irreducible representations, and one 3-dimensional irreducible representation. By a trick proposed in~\cite{ju}, we can quickly compute the determinant of the irreps using the inverse of the Cartan matrix. The result is as follows. The three 1-dimensional irreps $\rho_1, \rho'_1, \rho''_1$ have determinant, $1, \omega, \omega^2$, where $\omega$ is the third root of unity. The three 2-dimensional irreps $\rho_2, \rho'_2, \rho''_2$ have determinant $1,\omega,\omega^2$. The 3-dimensional irrep $\rho_3$ has determinant 1. There are seven conjugacy classes labled by $I, -I, \beta, \gamma, \gamma^2, \gamma^4, \gamma^5$, and the character table for $E_6$ is as follows~\cite{e6}. The number in the parenthesis in the first line represents the order of any element in the conjugacy class. 

\begin{table}[htbp]
\centering
\begin{tabular}{|c|c|c|c|c|c|c|c|}
\hline
\ & $I (1)$ & $-I (1)$ & $\beta (4)$ & $\gamma (6)$ & $\gamma^2 (3)$  &  $\gamma^4 (3)$ &  $\gamma^5 (6)$ \\ \hline
$\rho_1$ & 1 & 1 & 1 & 1 & 1  & 1 & 1 \\ \hline
$\rho_2$ & 2 & -2 & 0 & 1 & -1  & -1 & 1 \\ \hline
$\rho_3$ & 3 & 3 & -1 & 0 & 0  &  0 &  0 \\ \hline
$\rho'_2$ & 2 & -2 & 0 & $\omega$ & $-\omega$  &  $-\omega^2$ &  $\omega^2$ \\ \hline
$\rho''_2$ & 2 & -2 & 0 & $\omega^2$ & $-\omega^2$  &  $-\omega$ &  $\omega$ \\ \hline
$\rho'_1$ & 1 & 1 & 1 & $\omega$ & $\omega$  &  $\omega^2$ &  $\omega^2$ \\ \hline
$\rho''_1$ & 1 & 1 & 1 & $\omega^2$ & $\omega^2$  &  $\omega$ &  $\omega$ \\ \hline
\end{tabular}
\caption{Character table for the group $E_6$, retrieved from~\cite{e6}. Here, $\omega$ is the third root of unity. The number in the parenthesis represents the order of the conjugacy class.\label{table:e6character}}
\end{table}

We pick the two generators to sit in the $\beta$ and the $\gamma$ conjugacy classes and without any confusion we use the same letters to denote the generators. The defining two-dimensional geometric actions by these two generators are
\begin{equation}
\label{eq:betagamma}
    \beta = \begin{pmatrix}
        0 & i \\
        i & 0
    \end{pmatrix} \quad 
    \gamma = \frac{1}{\sqrt{2}}\begin{pmatrix}
        \epsilon & \epsilon^3 \\
        \epsilon & \epsilon^7
    \end{pmatrix}
\end{equation}
where $\epsilon$ is the eighth root of unity. To see that these two generators indeed generate the whole group, we can look at the quarternionic representation of $E_6$. The 24 group elements can be represented by the following unit quarternions.
\begin{itemize}
    \item 1 element of order 1: 1.
    \item 1 element of order 2: $-1$.
    \item 6 elements of order 4: $\pm i, \pm j, \pm k$.
    \item 8 elements of order 6: $(1\pm i \pm j \pm k)/2$.
    \item 8 elements of order 6: $(-1\pm i \pm j \pm k)/2$.
\end{itemize}
To make contact with Table~\ref{table:e6character}, we note that the first three lines each forms a conjugacy class, and each of the last two lines (elements of order 6) splits into two conjugacy classes. We rewrite the above using the language of conjugacy classes:
\begin{itemize}
    \item $I$: 1 element of order 1: 1.
    \item $-I$: 1 element of order 2: $-1$.
    \item $\beta$: 6 elements of order 4: $\pm i, \pm j, \pm k$.
    \item $\gamma$: 4 elements of order 6: $(1+i+j+k)/2$, with even numbers of sign flips for $i,j,k$.
    \item $\gamma^5$: 4 elements of order 6: $(1-i+j+k)/2$, with odd numbers of sign flips for $i,j,k$.
    \item $\gamma^2$: 4 elements of order 3: $(-1+i+j+k)/2$, with even numbers of sign flips for $i,j,k$.
    \item $\gamma^4$: 4 elements of order 3: $(-1-i+j+k)/2$, with odd numbers of sign flips for $i,j,k$.
\end{itemize}

One can check that, for example, $i$ and $(1+i+j+k)/2$ can generate all elements in the group. In the following, we shall therefore take $\beta, \gamma$ in equation~\eqref{eq:betagamma} as the generators, and we would like to find the $SU(2)$ Wilson lines representing these two elements. 

How many ground state Wilson lines are on $S^3/E_6$? The generating function for computing the number of $SU(2)$ representations of $E_6$, defined in the same fashion as the $\text{Dic}_k$ case as in~\eqref{eq:generating}, is~\cite{ganorandju} 
\begin{align*}
    \Phi_{E_6} &=\frac{1}{3} \left(\frac{1}{(1-z)^3(1-z^2)^3(1-z^3)}+\frac{2}{(1-z^6)(1-z^3)^2}\right) \\
    &=1+z+3z^2+...
\end{align*}
which suggests that there are 3 $SU(2)$ ground state Wilson lines. From the character table and from our previous discussion on the determinant of the representations, it is easy to see that the ground state Wilson lines are
\begin{equation}
    \rho_1\oplus \rho_1, \quad \rho_2, \quad \rho'_1 \oplus\rho''_1
\end{equation}
where $\rho_1\oplus \rho_1$ is the trivial Wilson line. In the following, we discuss the trivial Wilson line only. The only relevant action for the trivial Wilson line is the geometric action. Start with a $j_2=0$ field $\Phi$. The descendant $\partial_{++}^{n_1}\partial_{+-}^{n_2}\Phi$ transforms under the two generator $\beta$ as (see equation~\eqref{eq:betagamma})
\begin{equation}
    \beta(\partial_{++}^{n_1}\partial_{+-}^{n_2}\Phi) = i^{n_1+n_2} \partial_{++}^{n_2}\partial_{+-}^{n_1}\Phi
\end{equation}

What makes the index on $S^3/E_6$ so hard to compute compared with the A- and the D-singularity case is the action of $\gamma$:
\begin{align*}
    \gamma(\partial_{++}^{n_1}\partial_{+-}^{n_2}\Phi) &= \frac{1}{2^{(n_1+n_2)/2}} (\epsilon\partial_{++}+\epsilon^3 \partial_{+-})^{n_1}(\epsilon\partial_{++}-\epsilon^3 \partial_{+-})^{n_2} \Phi \\ 
    &= \frac{1}{2^{(n_1+n_2)/2}} \sum_{l=0}^{n_1}\sum_{m=0}^{n_2} {n_1 \choose l}{n_2\choose m} (-1)^{n_2-m}  \epsilon^{3n_1+3n_2-2m-2l} \partial^{m+l}_{++} \partial^{n_1+n_2-l-m}_{+-} \Phi 
\end{align*}

The equation above says that, if we start with an operator $\partial^n_{++} \Phi$, then the action of $\gamma$ would generate a sum of terms
\[
\partial^{l}_{++}\partial^{m}_{+-}  \Phi
\]
such that $l+m=n$. Therefore, a good ansats for an operator so that it is potentially invariant under $\gamma$ is
\begin{equation}
    \sum_{l+m=n}C^{n}_{lm}\partial^l_{++}\partial^m_{+-} \Phi
\end{equation}
for a given $n$. Using this ansatz, one can impose the constraints from $\beta$ and $\gamma$ to determine what kind of $C^n_{lm}$ is nonzero. We leave the detailed calculation for a future work and encourage the reader to work on this as well. Here, we conjecture the result based on the $\textbf{Z}_k$ and $\text{Dic}_k$ results we found earlier (equation~\eqref{eq:wilsonsolutionzk} and equation~\eqref{eq:wilsonsolutiondick}). Both equations show that the Casimir energy has to do with the \textit{order} of the discrete group $G$ as
\[
E = \frac{2}{3 \abs{G}}
\]
where $\abs{G} = k$ for $\mathbb{Z}_k$ and $4k$ for $\text{Dic}_k$. In the case of $E_6$, we have that $\abs{E_6}=24$, so we conjecture that the degenerate Casimir energy in the current regularization scheme is
\[
E  = \frac{1}{36}
\]

\section{Comparison with $\mathcal{N}=2$ Supersymmetry}
\label{sec:comp1}

We mentioned earlier that the exact degeneracy in the ground state supersymmetric Casimir energy is a special feature for the $\mathcal{N}=4$ theory and does not hold for less supersymmetric theories. In this section, we compute the supersymmetric Casimir energy for the $\mathcal{N}=2$ $SU(2)$ super Yang-Mills theory on $S^3/\text{Dic}_k$ to give support to this claim. The R-symmetry for $\mathcal{N}=2$ supersymmetry is $U(2)$, which can be decomposed into an $SU(2)$ part with quantum number $R$ and a $U(1)$ part with quantum number $r$. We pick the same $Q$ as in the $\mathcal{N}=4$ theory to define the index. The anticommutator in the $\mathcal{N}=2$ theory is
\begin{equation}
\label{eq:delta}
    2\{Q^\dagger,Q\} = D- 2J_1 - R - \frac{r}{2}
\end{equation}

The BPS operators annihilated by the above anticommutator are listed in Table~\ref{table:n=2bpstable}. 

\begin{table}[htbp]
\centering
\begin{tabular}{|c|c|}
    \hline
    Letter & $(-1)^F [E, j_1, j_2]$  \\
    \hline
    X & $[1,0,0]$ \\
    \hline
    $\psi_{1+}$ & $-[3/2,1/2,0]$ \\
    \hline
    $F_{++}$ & $[2,1,0]$ \\
    \hline
    $\partial_{++}\psi^2_{-} +\partial_{+-}\psi^2_{+}=0$ & $[5/2,1/2,0]$ \\
    \hline
    $\bar{\psi}_{1\pm}$ & $-[3/2,0,\pm 1/2]$ \\
    \hline
    $\bar{q}$ & $[1,0,0]$ \\
    \hline
    $\psi_{+}$ & $-[3/2, 1/2,0]$\\
    \hline
    $\partial_{+\pm}$ & $[1,1/2,\pm 1/2]$ \\
    \hline
\end{tabular}
\caption{A list of operators in $N=2$ supersymmetric theories that satisfy the BPS condition and their $E, j_1, j_2$ quantum numbers. The first operators along with the derivative operators are relevant for the vector multiplet, whereas $\bar{q}$ and $\psi_{+}$ along with the derivative operators are relavant for the hypermultiplet. The $R, r$ quantum numbers are omitted and can be found in the appendix of~\cite{benini}. \label{table:n=2bpstable}}
\end{table}

In the $\mathcal{N}=2$ $SU(2)$ SYM theory, one can have hypermultiplets, vector multiplets, or a combination of both so that the ``quarks'' are charged under some flavor symmetry. One can also consider the class-S theory~\cite{gaiotto} in which the topology of the Riemann surface determines the matter and gauge content of the theory. In the following subsections, we compute the supersymmetric Casimir energy for 1) the $SU(2)$ vector multiplet, 2) the $SU(2)$ hypermultiplet which transforms in the fundamental, and 3) the conformal $N_f=4$ theory. We also comment on the general class-S theory. 

One thing to note is that the $\mathcal{N}=2$ vector multiplet (or the hypermultiplet) by itself does not lead to a conformal field theory, and so we cannot interpret the calculation as computing the conformal dimension of the ground state Wilson line operators. Also, the state-operator correspondence will be lost and we cannot interpret each term in the Boltzmann sum as corresponding to a particular state of the theory. However, it is nevertheless okay to compute the index away from the conformal point by using the BPS spectrum at a conformal point.

\subsection{$\mathcal{N}=2$ Vector Multiplet}
The $\mathcal{N}=2$ vector multiplet transforms in the adjoint of $SU(2)$, so in computing the contribution of the constrained sum $t^{3n_13+n_2}$ to the single letter index, we can simply re-use the constrained sum we obtained for the $\mathcal{N}=4$ theory. The only difference is the BPS field content\footnote{For example, in the $\mathcal{N}=4$ theory there are six real scalar field which leads to three holomorphic combinations, explaining the letters $X,Y,Z$ in Table~\ref{table:n=4bpstable}. In the $\mathcal{N}=2$ theory there are two real scalar fields in the vector multiplet, which leads to only one holomorphic combination. This is why in Table~\ref{table:n=2bpstable} there is only one BPS scalar field as indicated by the letter ``X''.}. To compute the single letter index for the four trivial abelian Wilson lines, we can use the Boltzmann sum $F^{\text{abelian}}_{\text{Dic}_k, j_2=0}$ for $j_2=0$ BPS fields from equation~\eqref{eq:appendixfdickabelianj0} and $F^{\text{abelian}}_{\text{Dic}_k, j_2=\pm 1/2}$ for $j_2=\pm 1/2$ BPS fields as before:
\begin{equation}
    \hat{I}^{\text{vec},\text{abelian}}_{\text{Dic}_k} =  F^{\text{abelian}}_{\text{Dic}_k, j_2=0} (t^2-t^4+2t^6) - F^{\text{abelian}}_{\text{Dic}_k, j_2=\pm 1/2} t^3
\end{equation}
where $t^2-t^4+2t^6$ is the Boltzmann weight for the $j_2=0$ BPS base fields and $-t^3$ is the Boltzmanna weight for the $j_2=\pm 1/2$ base fields as in Table~\ref{table:n=2bpstable}. Expanding to first order in $\beta$, on obtains
\begin{equation}
\label{eq:n=2vectorablianI}
    \hat{I}^{\text{vec},\text{abelian}}_{\text{Dic}_k} = -\frac{2}{3 k \beta} + 3 + \frac{-2-27k-9k^2}{18k} \beta  + O(\beta)
\end{equation}

For the nonabelian Wilson line ground states labeled by $n=1,3,...,k-1$, we can use $F^{n}_{\text{Dic}_k, j_2=0}$ from~\eqref{eq:appendixdicf0} and $F^{n}_{\text{Dic}_k, j_2=\pm 1/2}$ from~\eqref{eq:appendixdicf1} to compute the single letter index
\begin{equation}
    \hat{I}^{\text{vec},n}_{\text{Dic}_k} = F^{n}_{\text{Dic}_k, j_2=0} ( t^2-t^4 + 2t^6) -F^{n}_{\text{Dic}_k, j_2=\pm 1/2} t^3
\end{equation}

Expanding to first order in $\beta$, we have
\begin{equation}
\label{eq:n=2vectornonablianI}
    \hat{I}^{\text{vec},n}_{\text{Dic}_k}= -\frac{2}{3k\beta} + \frac{-2+9k-9k^2+36nk-36n^2}{18k} \beta + O(\beta^2)
\end{equation}

As anticipated at the end of section~\ref{sec:a} and as can be seen from equation~\eqref{eq:n=2vectorablianI} and equation~\eqref{eq:n=2vectornonablianI}, the single letter index here (for both the abelian ground states and the nonabelian ground states) has a divergent term proportional to $1/\beta$, which is not present in the $\mathcal{N}=4$ SUSY case. We can read off the supersymmetric Casimir energy of the abelian ground states and for the nonabelian ground states labled by $n$:
\begin{align}
\label{eq:n=2vectorcasimirabelian}
    E^{\text{vec},\text{abelian}}_{\text{Dic}_k} &= \frac{2+27k+9k^2}{36k} \\
\label{eq:n=2vectorcasimir}
    E^{\text{vec},n}_{\text{Dic}_k} &= \frac{2-9k+9k^2-36nk+36n^2}{36k}
\end{align}
which shows that the degeneracy is partly lifted as the supersymmetric Casimir energy depends on $n$ for the nonabelian ground states.

\subsection{$\mathcal{N}=2$ Hypermultiplet}
The $\mathcal{N}=2$ hypermultiplet transforms in the fundamental of $SU(2)$. The hypermultiplet fields transform differently under the $SU(2)$ Wilson line from the vector multiplet fields, so we need to recompute the Boltzmann sum $\sum t^{3n_1+3n_2}$ for the derivative operators. One simplification here is that neither of the hypermultiplet base fields, $\bar{q}$ and $\psi_+$, has $j_2=\pm 1/2$, so we only need to consider one Boltzmann sum for the $j_2=0$ fields.For the $k/2$ nonabelian Wilson line ground states, a generic BPS operator is given by the linear combination
\begin{equation}
    \Psi = \sum C_{n_1n_2 \mu} \partial^{n_1}_{++} \partial^{n_2}_{+-} \Phi_\mu
\end{equation}
where $\mu=\pm 1$ is the $SU(2)$ fundamental index. We again demand two constraints for each Wilson line labeled by $n=1,3,5...,k-1$ (see equation~\eqref{eq:wilsondicksol5}):
\begin{align}
    \sum C_{n_1n_2 \mu}\exp(\pi i (n_1-n_2)/k) \partial^{n_1}_{++} \partial^{n_2}_{+-} \Phi_\mu &=\sum C_{n_1n_2 \mu}\exp(\pi i n\mu /k) \partial^{n_1}_{++} \partial^{n_2}_{+-} \Phi_\mu \\
    \sum C_{n_1n_2 \mu}\exp(\pi i n_2) \partial^{n_2}_{++} \partial^{n_1}_{+-} \Phi_\mu &=\sum C_{n_1n_2 \mu} (-1)^{(\mu+1)/2} \partial^{n_1}_{++} \partial^{n_2}_{+-} \Phi_{-\mu}
\end{align}

They imply
\begin{align}
n_1-n_2 -n\mu  &= 0\mod{2k} \\
C_{n_2n_1,-\mu} \exp(\pi in_1) &= (-1)^{(\mu+1)/2} C_{n_1n_2\mu}
\end{align}

The second equation tells us that we can just set $\mu=1$ and sum over all $(n_1, n_2)$ pairs such that the first equation is satisfied. We do not have to worry about the $\mu=-1$ case. The constrained Boltzmann sum therefore gives
\begin{equation}
    \sum t^{3n_1+3n_2} = \frac{t^{3n}+t^{6k-3n}}{(1-t^6)(1-t^{6k})}
\end{equation}

From Table~\ref{table:n=2bpstable}, the Boltzamnn weight for the scalar BPS field $\bar{q}$ is $t^2$ and that for the fermion BPS field $\psi_+$ is $-t^4$. So overall the single letter index for the $SU(2)$ hypermultiplet on $S^3/\text{Dic}_k$ for the $n$th nonabelian ground state is
\begin{equation}
    \hat{I}^{\text{hyper},n}_{\text{Dic}_k} = \frac{t^{3n}+t^{6k-3n}}{(1-t^6)(1-t^{6k})} (t^2-t^4)
\end{equation}

Expanding to first order in $\beta$, we obtain
\begin{equation}
    \hat{I}^{\text{hyper},n}_{\text{Dic}_k} = \frac{2}{9k\beta} + \frac{-8+18k^2-54nk+27n^2}{108k} \beta + O(\beta^2)
\end{equation}
which implies a supersymmetric Casimir energy of 
\begin{equation}
\label{eq:n=2hypercasimir}
    E^{\text{hyper},n}_{\text{Dic}_k} = \frac{8-18k^2+54nk-27n^2}{216k}
\end{equation}

Similarly, one can compute the single letter index and the supersymmetric Casimir energy for the four abelian ground states. The results are
\begin{align}
\label{eq:n=2hypercasimirabelian1}
    E^{\text{hyper, abelian},1}_{\text{Dic}_k} &=\frac{4-27k-9k^2}{36k} \\
    \label{eq:n=2hypercasimirabelian2}
    E^{\text{hyper, abelian},2}_{\text{Dic}_k} &=\frac{4+27k-9k^2}{36k} \\
    \label{eq:n=2hypercasimirabelian3}
    E^{\text{hyper, abelian},3}_{\text{Dic}_k} &=\frac{8+9k^2}{216k} \\
    \label{eq:n=2hypercasimirabelian4}
    E^{\text{hyper, abelian},4}_{\text{Dic}_k} &=\frac{8+9k^2}{216k} 
\end{align}

From the above equations, we see that just as the vector multiplet case~\eqref{eq:n=2vectorcasimir}, for the hypermultiplet the degeneracy is also partially lifted and a discrete quadratic potential generated for the nonabelian ground states.  

\subsection{$\mathcal{N}=2$ $N_f=4$ Theory}
In this section we consider the $\mathcal{N}=2$ $N_f=4$ $SU(2)$ theory where $N_f$ indicates the number of flavors. This means that the theory contains one $SU(2)$ vector multiplets and four $SU(2)$ hypermultiplets. This theory is special because it is conformal. One way to see this is to compute the $\beta$ function explicitly. Another way to see this is to use the D-brane construction~\cite{wittenbranes} where we put two NS5 branes separeted in the 6 direction and two D4 branes in between the NS5 branes. The worldvolume of the NS5 brane is along the 012345 direction, while that of the D4 brane is along the 01236 direction. The D4 branes create dimples at where they intersect the NS5 branes, bending the NS5 branes toward the D4 branes. This bending is interpreted as the running coupling constant. To make the NS5 branes straight (no running coupling constant), one can attach two semi-infinite D4 branes to the left of the first NS5 brane and to the right of the second NS5 brane. These four semi-infinite D4 branes create the four hypermultiplets in the $N_f=4$ theory. 

To find the supersymmetric Casimir energy of this theory, we simply have to add four times the hypermultiplet Casimir energy as in~\eqref{eq:n=2hypercasimir} to the vector multiplet Casimir energy as in~\eqref{eq:n=2vectorcasimir}. Therefore, the supersymmetric Casimir energy for the nonabelian Wilson line ground states for this $N_f=4$ theory is
\begin{equation}
    E^{N_f=4,n}_{\text{Dic}_k} = \frac{11}{54k} - \frac{1}{4} - \frac{1}{12}k + \frac{n^2}{2k}, \quad n=1,3,5...,k-1
\end{equation}

For the four abelian Wilson line ground states, we have (the trivial Wilson line ground state is labled by 1)
\begin{align}
    \label{eq:nf=4trivial}
    E^{N_f=4,\text{abelian}, 1}_{\text{Dic}_k} &= \frac{11}{54k} - \frac{1}{4} - \frac{1}{12}k   \\
    E^{N_f=4,\text{abelian}, 2}_{\text{Dic}_k} &= \frac{11}{54k} + \frac{7}{4} - \frac{1}{12}k   \\
    E^{N_f=4,\text{abelian}, 3}_{\text{Dic}_k} &= \frac{11}{54k} + \frac{3}{4} + \frac{7}{24}k   \\
    E^{N_f=4,\text{abelian}, 4}_{\text{Dic}_k} &= \frac{11}{54k} + \frac{3}{4} + \frac{7}{24}k 
\end{align}

From the five equations above, we see that degeneracy of the $N_f=4$ theory ground states is also partly lifted. The most important and surprising feature of the above equations is that the superconformal Casimir energy is minimized for the unique ground state: the trivial Wilson line ground state as in equation~\eqref{eq:nf=4trivial}. This is important for the following reason. In 2D CFT, the conformal dimension for operators is defined on $\mathbb{C}$. The Casimir energy for 2D operators is computed on a cylinder. The Casimir energy and the conformal dimension are related through the anomalous transformation property of the energy momentum tensor $T$ (which is not primary)
\begin{equation}
    z^2 T_{zz}  = T_{ww} + \frac{c}{24} 
\end{equation}
where $c$ is the central charge of the 2D CFT, $z$ is the coordinate of $\mathbb{C}$ and $w$ is the coordinate on the cylinder. This equation implies that there is a universal shift constant $c/24$ that must be added to the Casimir energy in order to obtain the conformal dimension of an operator. 

The computation of the supersymmetric Casimir energy that we are doing in this paper is analogous to the Casimir energy on the 2D cylinder. To obtain the conformal dimensions of the ground state Wilson lines, one should add a constant to each Casimir energy value. This constant is characterized by the 4D conformal anomaly. Since the conformal anomaly depends on the geometry only, each ground state Casimir energy must be shifted by the same constant to obtain the conformal dimension. Here, for the $N_f=4$ theory, we assume that the trivial Wilson line corresponds to the unit operator, which has a conformal dimension of 0. According to equation~\label{eq:nf=4trivial}, this shift constant is simply 
\begin{equation}
    \frac{11}{54k} - \frac{1}{4} - \frac{1}{12}k  
\end{equation}

Applying the shift to all ground state Wilson lines, one finds their conformal dimensions $D$ to be\footnote{We assume that the ground state Wilson lines do not belong to some $J_1$ multiplet, i.e they all have $J_1=0$, so that the supersymmetric Casimir energy $D+J_1$ is really the conformal dimension after some constant shift by the conformal anomaly.}:

\begin{align}
    D^{N_f=4,\text{abelian}, 1}_{\text{Dic}_k} &= 0 \\
    D^{N_f=4,\text{abelian}, 2}_{\text{Dic}_k} &= 2 \\
    D^{N_f=4,\text{abelian}, 3}_{\text{Dic}_k} &= 1+\frac{3}{8}k \\
    D^{N_f=4,\text{abelian}, 4}_{\text{Dic}_k} &= 1+\frac{3}{8}k \\
    D^{N_f=4,n}_{\text{Dic}_k} &=\frac{n^2}{2k}, \quad n=1,3,5,...,k-1  
\end{align}

In particular, the trivial Wilson line ground state, having conformal dimension 0, is the true ground state of this theory. All other classical ground states have positive conformal dimension. This result is significant because the $N_f=4$ conformal theory, being superconformal, has a unitarity bound, so all operators should have nonnegative conformal dimensions. It is impossible for nontrivial ground state Wilson lines to obtain a supersymmetric Casimir energy less than that of the trivial ground state Wilson line. On the other hand, this result is also surprising. The uninitiated might be tempted to conclude that the ground states for the conformal $SU(2)$ $N_f=4$ theory on $S^3/\Gamma$ ($\Gamma=\text{Dic}_k$) are simply Weyl-inequivalent homomorphisms from $\Gamma$ to $SU(2)$ as he would do for the case of $\mathcal{N}=4$ SUSY. Our detailed calculation in this section shows that for the $N_f=4$ theory there is only one unique ground state corresponding to the trivial Wilson line, unlike the case of $\mathcal{N}=4$ SUSY where all classically flat Wilson lines have the same conformal dimension 0 (after shifting by some anomaly constant). 

How should we make of this surprising result? We postpone this discussion to the end of the next subsection on class-S theory.

\subsection{Class-S Theory}
One way to derive the duality proposed in~\cite{ganorandju,vafawitten,dijkgraaf} is to consider a stack of $q$ M5 branes compactified on a torus $T^2$ (for this derivation, see section 9 of~\cite{ju}). The ADE singularity $\Gamma$ acts on $\mathbb{R}^4$, the rest of the world volume of the M5 branes. It was found that the dual Chern-Simons theory with level $q$ and gauge group given by $G(\Gamma)$ (recall that $G(\Gamma)$ is the gauge group McKay dual to $\Gamma$) lives on the torus $T^2$. The number of ground states on the Chern-Simons theory side is simply the number of level $q$ WZW conformal blocks~\cite{wittenjones}. This number is shown in~\cite{ganorandju,ju,vafawitten,dijkgraaf} to coincide with the number of flat connections on the $U(q)$ $N=4$ SYM theory\footnote{If the SYM theory has gauge group $SU(q)$ instead of $U(q)$, the dual Chern-Simons states are a restricted class of the WZW conformal blocks~\cite{ganorandju,ju}.} on $S^3/\Gamma$. 

A variant of this construction is to replace the $T^2$ by a genus $g$ Riemann surface. It is natural to expect that the ground state duality still holds, except that the theories on both sides of the duality are modified. 

\begin{itemize}
\item The SYM side: The supersymmetry is reduced from $\mathcal{N}=4$ to $\mathcal{N}=2$. The theory becomes the $\mathcal{N}=2$ class-S theory~\cite{gaiotto}. Let the number of M5 branes be two. The fundamental building block of the class-S theory in this case is the trinion theory. One way to visualize it is to imagine a vertex with three legs (or, in the blow-up limit, a pair of pants). The legs (either form the same vertex or different vertices) can be contracted to obtain a graph which corresponds to some degeneration limit of the Riemann surface. Each contracted leg yields a $\mathcal{N}=2$ vector multiplet, and hence a gauge field to create a Wilson line that wraps around the ADE singularity. 

\item The Chern-Simons side: The Chern-Simons theory is still the same as before with the same level $q$ and the same gauge group $G(\Gamma)$, except that it is quantized on the genus $g$ Riemann surface instead of on $T^2$. The fundamental building blocks of the ground states of this theory are the fusion rule coefficients $N_{ijk}$. Any genus-two Riemann surfaces can be constructed by splicing together pair-of-pants topologies. The three holes of the pants correspond to the indices $i,j,k$. The pants are contracted to form the Riemann surface, and the number of Chern-Simons ground states is simply a contraction of a series of fusion rule coefficients $N_{ijk}$.
\end{itemize}

Naively, if a degeneration limit of the Riemann surface of the $SU(2)$ class-S theory has $p$ internal lines, and if each vector field can create $h$ flat Wilson lines by wrapping around the ADE singularity, one would expect (without knowing the duality argument) the number of ground states to be $p^h$. This would have been true had we still been dealing with $\mathcal{N}=4$ theories; we showed earlier that exact degeneracy of supersymmetric Casimir energy does not occur for $\mathcal{N}=2$ theories. There has to be a truncation of ground states, and in the language of the duality this truncation comes about because not all $N_{ijk}$ are nonzero. A zero $N_{ijk}$ would imply an impossible combination of Wilson lines at a particular trinion vertex. Such a combination of Wilson lines will lead to a supersymmetric Casimir energy different from the one for the state involving only the trivial Wilson lines ($N_{111}=1$ where the subscript 1 corresponds to the identity element). The detail of this computation is fleshed out in an upcoming work~\cite{emil}.

Now, let us go back to interpreting the surprising result from the previous subsection that there is only one unique ground state for the $SU(2)$ $N_f=4$ theory. This theory can be obtained by putting two M5 branes on a four-punctured sphere. Therefore, the dimension of the ground states of the $SU(2)$ $N_f=4$ theory can be computed using the dual Chern-Simons theory on the four-punctured sphere as
\begin{equation}
    \sum_k N_{\tilde{i}\tilde{j} k} N^{k \tilde{l}\tilde{m}}
\end{equation}
where $\tilde{i},\tilde{j},\tilde{l},\tilde{m}$ are the four (in and out) states specified on the four punctures. We conjecture that the most natural choice for the in and the out states is to put them as the identity $1$, so that only $k=1$ contributes the sum, resulting in a 1-dimensional Hilbert space\footnote{We are thankful for O.~Ganor for bringing up this point of four-punctured sphere.}.

\section{Implications For S-Duality}
\label{sec:sduality}

S-duality maps the weak coupling regime of the $\mathcal{N}=4$ SYM theory with gauge group $G$ to the strong coupling regime of the same theory with the Langlands dual gauge group $\tilde{G}$~\cite{goddard}. It exchanges the electric degrees of freedom with the magnetic degrees of freedom and elementary particles with solitons. In particular, the Wilson lines are exchanged with the t' Hooft lines~\cite{thooft1,thooft2}. In our setting, S-duality turns the ground state Wilson lines wrapping the $\Gamma$ singularity into the ground state t' Hooft lines. Because the Wilson lines take values in $SU(2)$, the ground state t' Hooft lines will take values in the Langlands dual group $SU(2)/\mathbb{Z}_2 \approx SO(3)$.  S-duality makes two predictions here. First, the number of ground state t' Hooft lines must be equal to the number of ground state Wilson lines. In other words, the number of $SU(2)/\mathbb{Z}_2$ representations of $\Gamma$ (up to identification by conjugation and Weyl group) must be equal to the number of $SU(2)$ representations of $\Gamma$. Second, for a given $\Gamma$, the ground state t' Hooft lines must be exactly degenerate in the supersymmetric Casimir energy, just like their Wilson line counterparts. In addition to this degeneracy, the supersymmetric Casimir energy of the t' Hooft lines must be the same as their Wilson line counterparts. This is because S-duality maps the weakly-coupled electric ground state Hilbert space to the strongly-coupled magnetic ground state Hilbert space. 

In the following subsections, we check these two predictions explicitly for the A- and the D-singularity using the gauge group $SU(2)$. Since the supersymmetry Casimir energy $D+j_1$ is protected by supersymmetry, we can imagine doing an index calculation using the dual theory t' Hooft lines and dual theory BPS fields to compute the supersymmetry Casimir energy of the dual theory. 

\subsection{S-Duality on $S^3/\mathbb{Z}_k$}
\label{sec:sdualityzk}
As mentioned earlier, the dual t' Hooft line will take values in $SU(2)/\mathbb{Z}_2$, furnishing some representation of $SU(2)/\mathbb{Z}_2\to\mathbb{Z}_k$. In other words, let a t' Hooft line be $\tilde{g}(r)\in SU(2)/\mathbb{Z}_2$. It must satisfy the relation
\begin{equation}
\label{eq:thooftzk}
    \tilde{g}(r)^k = \pm \textbf{1}
\end{equation}
and we identify two t' Hooft lines $\tilde{g}(r)$ and $\tilde{g}'(r)$ if they differ by $-\textbf{1}$ (and by conjugation and Weyl group transformation). We would like to show that the number of $SU(2)/\mathbb{Z}_2\to\mathbb{Z}_k$ representations is the same as that of $SU(2)\to\mathbb{Z}_k$ representations. Before we prove this, we give a specific example for $k=2$. In this case, there are two Wilson lines
\begin{align*}
    g_0(r) &= \begin{pmatrix}
        1 & 0 \\
        0 & 1
    \end{pmatrix}  \\
    g_1(r) &= -\begin{pmatrix}
        1 & 0 \\
        0 & 1
    \end{pmatrix}
\end{align*}

However, in the t' Hooft line picture, these two solutions correspond to the same t' Hooft line because they differ by $-\textbf{1}$. A moment's thought reveals that there are indeed two t' Hooft lines:
\begin{align*}
    \tilde{g}_0(r) &= \begin{pmatrix}
        1 & 0 \\
        0 & 1
    \end{pmatrix}  \\
    \tilde{g}_1(r) &= \begin{pmatrix}
        i & 0 \\
        0 & -i
    \end{pmatrix}
\end{align*}

Although $\tilde{g}_1(r)$ does not square to $\textbf{1}$, it squares to $-\textbf{1}$ which is allowed by~\eqref{eq:thooftzk}. We now prove the general case. First, consider the case where $k$ is odd. In this case, the claim is that the t'Hooft lines take the same values as the Wilson lines
\begin{equation}
\label{eq:zksol}
    \tilde{g}_j(r) = \begin{pmatrix} 
    e^{2\pi ij/k} & 0 \\
    0 & e^{-2\pi ij/k} 
    \end{pmatrix}, \quad j=0,1,...,\floor*{\frac{k}{2}}
\end{equation}

It is easy to see that no two solutions are identified under conjugation or under the Weyl group. To see that no two solutions are identified by multiplication by $-\textbf{1}$, we assume the contrary and suppose that for some $p, q\in \mathbb{Z}$
\begin{equation}
    \tilde{g}_p(r) = -\tilde{g}_q(r)
\end{equation}
or
\begin{equation}
    e^{2\pi i(p-q)/k} = -1
\end{equation}

But this is impossible since, $k$ being an odd integer by assumption, $(p-q)/k$ can never be an odd mulitple of $1/2$. In addition to the solutions~\eqref{eq:zksol}, one can also have solutions of the form
\begin{equation}
    \tilde{g}'_j(r)=\begin{pmatrix}
    e^{\pi ij/k} & 0 \\
    0 & e^{-\pi ij/k}
    \end{pmatrix}, \quad j=1,3,...,k
\end{equation}
since their $k$th power is $-\textbf{1}$, satisfying equation~\eqref{eq:thooftzk}. However, these solutions are redundant: each of these solutions is identified with an old solution in equation~\eqref{eq:zksol}. To see this, consider $-1$ times $\tilde{g}'_j(r)$, which gives
\begin{equation}
    -\tilde{g}'_j(r)=\begin{pmatrix}
    e^{\pi i(j+k)/k} & 0 \\
    0 & e^{-\pi i(j+k)/k}
    \end{pmatrix}, \quad j=1,3,...,k
\end{equation}

By assumption, $k$ is odd, and so $j+k$ must be even. Solutions of this form are just the old solutions~\eqref{eq:zksol}. This concludes the proof that for odd $k$, the number of flat Wilson lines is the same as that of the t' Hooft lines. 
When $k$ is even, the solutions to the t' Hooft lines are of the form
\begin{equation}
\label{eq:zksol}
    \tilde{g}_j(r) = \begin{pmatrix} 
    e^{\pi ij/k} & 0 \\
    0 & e^{-\pi ij/k} 
    \end{pmatrix}, \quad j=0,1,...,k/2
\end{equation}

There are $k/2+1$ solutions, which agree with the number of Wilson line solutions. The proof is similar to the above, so we omit it. 

The next step is to compute the supersymmetric Casimir energy of the t' Hooft lines, which is expected to be the same as that of the Wilson lines. To do this, we imagine starting with the electric theory and dialing up the coupling constant $g_{YM}$ to a very large value. The $D+j_1$ value does not change in this process, since it is a protected quantity. Next, we use S-duality to go to the weakly coupled magnetic theory in which the elementary fields create/annihialte monopoles (as perceived in the electric frame). We assume that the spectrum of the BPS operators does not change, i.e. in the dual magnetic theory we have a similar BPS operator spectrum the same as Table~\ref{table:n=4bpstable}. We also assume that the elementary fields transform in the adjoint of the magnetic gauge group $SU(2)/\mathbb{Z}_2$. The last two assumptions imply that we can carry out the same index computation we did before for the $\mathbb{Z}_k$ case, except now using the t' Hooft line solutions. 

The case where $k$ is odd is easy because we proved earlier that the t' Hooft line solutions for $k$ odd are the same as the Wilson line solutions. Therefore, for odd $k$ the supersymmetric Casimir energy of the t' Hooft line ground states is the same as that of the Wilson line ground states given by equation~\eqref{eq:wilsonsolutionzk}. The novelty here is the $k$ even case. The ingredients that go into the single letter index for $k$ even are worked out in appendix~\ref{sec:appendixmagneticzk}. The single letter index $\tilde I^n$ for the $n$th t' Hooft line is  
\begin{align}
    \label{eq:thooftindexzk}
    \tilde{I}^0_{k} &= T^0_{\mathbb{Z}_k,j_2=0} (3t^2-3t^3+2t^6) - T^0_{\mathbb{Z}_k,j_2=\pm 1/2} t^3 \\
    \tilde{I}^n_{k} &= T^n_{\mathbb{Z}_k,j_2=0} (3t^2-3t^3+2t^6) - T^n_{\mathbb{Z}_k,j_2=\pm 1/2} t^3, \quad n\neq 0
\end{align}
where the $T$ functions are defined in equations~\eqref{eq:appendixzkmag1}~\eqref{eq:appendixzkmag2}~\eqref{eq:appendixzkmag3}~\eqref{eq:appendixzkmag4}. Expanding to first order in $\beta$, we have
\begin{align}
    \tilde{I}^0_{k} &= 3-\frac{4}{3k}\beta + O(\beta^2) \\
    \tilde{I}^n_{k} &= 1-\frac{4}{3k}\beta + O(\beta^2), \quad n\neq 0
\end{align}
which suggests that all $k/2+1$ t' Hooft line ground states have supersymmetric Casimir energy of $2/3k$, in agreement with the Wilson line ground state energy~\eqref{eq:wilsonsolutionzk}. This is a nontrivial statement of S-duality on the ground states for the exactly marginal $\mathcal{N}=4$ SYM theory. It does not in general hold for less supersymmetric theory.

\subsection{S-Duality on $S^3/\text{Dic}_k$}
\label{sec:sdualitydick}
Although the group $\text{Dic}_k$ is nonabelian, we expect that ground state S-duality holds on $S^3/\text{Dic}_k$ just as well since $\text{Dic}_k$ can be embedded in the same $SU(2)_L$ as $\mathbb{Z}_k$. To compare with the Wilson line results, we take $k$ to be even throughout this subsection. To find the t' Hooft lines, we need to solve for $r,s \in SU(2)/\mathbb{Z}_2$ so that the following equations are satisfied
\begin{equation}
    r^{2k} = \pm e, \quad s^2 =\pm r^k, \quad s^{-1}rs=\pm r^{-1}
\end{equation}
for any of the eight choices of the $\pm$ sign combination. Because there are two generators and because the group is nonabelian, the t' Hooft line solutions for the $\text{Dic}_k$ case are not as easy to find as the abelian $\mathbb{Z}_k$ case. For example, we found in section~\ref{sec:d} that there are $4+k/2$ Wilson line ground states. Under the $-1$ identification, the four abelian solutions~\eqref{eq:wilsondicksol1}~\eqref{eq:wilsondicksol2}~\eqref{eq:wilsondicksol3}~\eqref{eq:wilsondicksol4} are actually one and the same t' Hooft line solution. In addition, some of the nonabelian solutions are also identified under $-1$. As shown in appendix~\ref{sec:appendixmagneticdick}, there are indeed $4+k/2$ t' Hooft line solutions, in agreement with the Wilson line result. However, the structure of the t' Hooft line solutions is very different from that of the Wilson line solutions. For the Wilson line case, there are four universal solutions as in equations ~\eqref{eq:wilsondicksol1}~\eqref{eq:wilsondicksol2}~\eqref{eq:wilsondicksol3}~\eqref{eq:wilsondicksol4} and $k/2$ solutions having $k$ dependence as in equation~\eqref{eq:wilsondicksol5}. For the t' Hooft line solutions, there are five universal solutions and $k/2-1$ solutions that have a $k$ dependence\footnote{For the base case $\text{Dic}_2$, the only solutions are the five universal solutions.}. The former will be labeled by $\tilde{g}_j, j=1,...,5$, and the latter by $\tilde{g}'_n$, $n=2,4,...,k-2$:
\begin{align}
\label{eq:thooftdicksol1}
    \tilde{g}_1(r)&=\begin{pmatrix}
        1 & 0 \\
        0 & 1
    \end{pmatrix} \quad
    \tilde{g}_1(s)=\begin{pmatrix}
        1 & 0 \\
        0 & 1
    \end{pmatrix} \\
    \label{eq:thooftdicksol2}
    \tilde{g}_2(r)&=\begin{pmatrix}
        1 & 0 \\
        0 & 1
    \end{pmatrix} \quad
    \tilde{g}_2(s)=\begin{pmatrix}
        i & 0 \\
        0 & -i
    \end{pmatrix} \\
    \label{eq:thooftdicksol3}
    \tilde{g}_3(r)&=\begin{pmatrix}
        i & 0 \\
        0 & -i
    \end{pmatrix} \quad
    \tilde{g}_3(s)=\begin{pmatrix}
        1 & 0 \\
        0 & 1
    \end{pmatrix}\\
    \label{eq:thooftdicksol4}
    \tilde{g}_4(r)&=\begin{pmatrix}
        i & 0 \\
        0 & -i
    \end{pmatrix} \quad
    \tilde{g}_4(s)=\begin{pmatrix}
        i & 0 \\
        0 & -i
    \end{pmatrix}\\
    \label{eq:thooftdicksol5}
    \tilde{g}_5(r)&=\begin{pmatrix}
        i & 0 \\
        0 & -i
    \end{pmatrix} \quad
    \tilde{g}_4(s)=\begin{pmatrix}
        0 & 1 \\
        -1 & 0
    \end{pmatrix}
\end{align}

\begin{align}
\label{eq:thooftdicksol6}
\tilde{g}_n'(r)&=\begin{pmatrix}
\exp(n\pi i/2k) & 0 \\
0 & \exp(-n\pi i/2k) 
\end{pmatrix} \quad n=2,4,6,...,k-2 \\
\tilde{g}'_n(s)&=
\begin{pmatrix}
0 & 1 \\
-1 & 0
\end{pmatrix} \quad n=2,4,6,...,k-2 \nonumber
\end{align}

Having shown that the number of ground state t' Hooft lines is the same as that of the ground state Wilson lines, we now show that they also have the same supersymmetric Casimir energy as predicted by S-duality. Just as the $\mathbb{Z}_k$ case, the $\text{Dic}_k$ indices all have the same structure:
\begin{equation}
\label{eq:thooftdickindex}
    \tilde{I}^n_{\text{Dic}_k} = T^n_{\text{Dic}_k,j_2=0} (3t^2-3t^3+2t^6) - T^n_{\text{Dic}_k,j_2=\pm 1/2} t^3
\end{equation}

The $T$ functions can be calculated using the approach in appendix~\ref{sec:appendix2} and are tabulated in appendix~\ref{sec:appendixmagneticdick}. Using the result from the appendix, we can compute the single letter indices and expand to first order in $\beta$. The result is
\begin{align}
    \tilde{I}^1_{\text{Dic}_k} &= 3-\frac{\beta}{3k}\\
    \tilde{I}^2_{\text{Dic}_k} &= 1-\frac{\beta}{3k}\\
    \tilde{I}^3_{\text{Dic}_k} &= 1-\frac{\beta}{3k}\\
    \tilde{I}^4_{\text{Dic}_k} &= 1-\frac{\beta}{3k}\\
    \tilde{I}^5_{\text{Dic}_k} &= -\frac{\beta}{3k}\\
    \tilde{I}'^n_{\text{Dic}_k} &= -\frac{\beta}{3k}
\end{align}
where the first five lines $\tilde{I}^j_{\text{Dic}_k}$ are the single letter indices for the five universal solutions and the last line $\tilde{I}'^n_{\text{Dic}_k}$ is for the $k/2-1$ solutions, with the superscript $n=2,4,...,k-2$. Since the above indices all have the same first order term, all t' Hooft line ground states have the same supersymmetric Casimir energy $1/6k$ for a given $k$. Comparing this result with the Wilson line result~\eqref{eq:diccasimir}, we find that both the Wilson line ground states and the t' Hooft line ground states have the same supersymmetric Casimir energy $1/6k$, agreeing with the prediction of S-duality.

\section{Conclusion}
In this work, We find that, for the $\mathcal{N}=4$ $SU(2)$ SYM theory, the Wilson line ground states on $S^3/\Gamma$ where $\Gamma = \mathbb{Z}_k$ (lens space) or $\Gamma=\text{Dic}_k$ have the same supersymmetric Casimir energy. This result can be viewed as a one-loop test of the duality~\cite{ganorandju}\cite{vafawitten}\cite{dijkgraaf} that relates the ground states of $N=4$ $U(q)$ (or $SU(q)$) SYM on $S^3/\Gamma$ to the ground states (or a subspace of the ground states) of the level $q$ Chern-Simons theory with the McKay dual gauge group $G(\Gamma)$. Such degeneracy in the ground state supersymmetric Casimir energy is not found in theories with fewer than four supercharges. We showed this using the example of $\mathcal{N}=2$ supersymmetry and briefly mentioned the role of supersymmetry Casimir energy in class-S theory~\cite{emil}. In fact, for the conformal $N_f=4$ $SU(2)$ theory, our result shows that there is only one true ground state: the one involving the trivial Wilson line. This is surprising at first, but can be potentially explained by looking at the Hilbert space dimension of the corresponding Chern-Simons theory on a four-punctured sphere with all the in and the out states in the identity representation.

We also find that the number of t' Hooft line ground states equals that of the Wilson line ground states, and that the supersymmetry Casimir energy of the t' Hooft line ground states is the same as that of the Wilson line ground states. This provides yet another check of the prediction of S-duality. 

Although the numerics done in this paper assumes that the SYM theory has $SU(2)$ gauge group, we have good faith to conjecture (based on the well-motivated duality argument in~\cite{ganorandju,ju,vafawitten,dijkgraaf,nakajima}) that the degeneracy in supersymmetry Casimir energy in both the Wilson line sector and the t' Hooft line sector holds for all $SU(q)$ and $U(q)$ gauge group on $S^3/\Gamma$ where $\Gamma$ can be any of the ADE singularities. As discussed in~\cite{ganorandju,ju}, we remain ignorant about other classical gauge groups such as $SO(2q)$, $SO(2q+1)$, and $Sp(2q)$, and we leave the study of them as an open question.

\acknowledgments

The author would like to thank Emil Albrychiewicz, Andrés Franco Valiente, and Ori Ganor for useful discussions. The author also thanks Yasunori Nomura for his discussion on dilaton stabalization and Vera Serganova for her discussion on superalgebra.

\appendix
\section{$\mathbb{Z}_k$ Index Calculation}
\label{sec:appendix1}
In this appendix we work out the contribution of the $j_2=\pm 1/2$ fields to the index for $\mathbb{Z}_k$ to complement the $j_2=0$ result in section~\ref{sec:a}. Let $l=-1,0,1$ be the $SU(2)$ adjoint index, and $\mu=-1,1$ the doublet index. The latter responds under the geometric action $r$ as
\begin{equation}
    r\Phi_{\mu l} = \exp(2\pi i\mu/k) \Phi_{\mu l}
\end{equation}

The $n$th holonomy is $\exp(2\pi ni/k)$, where $n$ ranges from $0$ to $\floor*{k/2}$ in integer steps. Under the adjoint action of the $n$th holonomy $g_n(r)$, the field transform as
\begin{equation}
    g_n(r) (\Phi_{\mu 0}, \Phi_{\mu 1},\Phi_{\mu,-1}) g^{-1}_n(r) = (\Phi_{\mu 0}, \exp(4\pi in/k)\Phi_{\mu 1},\exp(-4\pi in/k)\Phi_{\mu,-1})
\end{equation}
as explained in section~\ref{sec:a}.

A general single letter operator can be written as 
\begin{equation}
    \Psi' = \sum C_{n_1n_2\mu l}  \partial_{++}^{n_1} \partial_{+-}^{n_2} \Phi_{\mu l}
\end{equation}

Because there is only one generator $r$ for the group $\mathbb{Z}_k$, there is only one constraint the operator $\Psi'$ needs to satisfy. It is given by equation~\eqref{eq:constraint} and reads
\begin{equation}
    \sum \exp(2\pi i(n_1-n_2+\mu)/k) C_{n_1n_2\mu l}  \partial_{++}^{n_1} \partial_{+-}^{n_2} \Phi_{\mu l} = \sum \exp(4\pi inl/k) C_{n_1n_2\mu l}  \partial_{++}^{n_1} \partial_{+-}^{n_2} \Phi_{\mu l}
\end{equation}

For $C_{n_1n_2\mu l}$ to be nonvanishing, one must have
\begin{equation}
\label{eq:appendixzk1}
    n_1-n_2 +\mu -2nl =0 \mod{k}
\end{equation}

Since $\mu$ can take values in $\{-1,1\}$ and $l$ in $\{-1,0,1\}$, there are six cases we need to consider. However, by the symmetry of equation~\eqref{eq:appendixzk1}, we only need to consider the case $\{l=0, \mu=-1\}$, $\{l=1,\mu=-1\}$, and $\{l=1,\mu=1\}$, since the rest of cases, $\{l=0,\mu=1\}$, $\{l=-1,\mu=1\}$, and $\{l=-1,\mu=-1\}$ give the same contribution to the single letter index as the previous three cases, respectively. Let us first assume that $k$ is even. 

\textbf{Case} $\{l=0,\mu=-1\}$. In this case we need to sum over all $(n_1,n_2)$ pairs in
\begin{equation}
    \sum t^{3n_1+3n_2}
\end{equation}
such that equation~\eqref{eq:appendixzk1} is satisfied for $l=0, \mu=1$, or
\begin{equation}
    n_1-n_2 -1=0\mod{k}
\end{equation}
The constrained sum gives
\begin{equation}
    \sum t^{3n_1+3n_2} = \frac{t^{3}+t^{3k-3}}{(1-t^6)(1-t^{3k})}
\end{equation}

\textbf{Case} $\{l=1,\mu=-1\}$. In this case we need to sum over all $(n_1,n_2)$ pairs in
\begin{equation}
    \sum t^{3n_1+3n_2}
\end{equation}
such that they satisfy the equation
\begin{equation}
    n_1-n_2 -1 - 2n =0\mod{k}
\end{equation}

If $n=0$ or $n=k/2$ (since we assumed $k$ even), the constrained sum is
\begin{equation}
    \sum t^{3n_1+3n_2} = \frac{t^3+t^{3k-3}}{(1-t^6)(1-t^{3k})}
\end{equation}

Otherwise, we have
\begin{equation}
    \sum t^{3n_1+3n_2} = \frac{t^{6n+3}+t^{3k-6n-3}}{(1-t^6)(1-t^{3k})}
\end{equation}

\textbf{Case} $\{l=-1,\mu=-1\}$. In this case we need to sum over all $(n_1,n_2)$ pairs in
\begin{equation}
    \sum t^{3n_1+3n_2}
\end{equation}
such that they satisfy the equation
\begin{equation}
    n_1-n_2 -1 + 2n =0\mod{k}
\end{equation}

If $n=0$ or $n=k/2$ (since we assumed $k$ even), the constrained sum is
\begin{equation}
    \sum t^{3n_1+3n_2} = \frac{t^3+t^{3k-3}}{(1-t^6)(1-t^{3k})}
\end{equation}

Otherwise, we have
\begin{equation}
    \sum t^{3n_1+3n_2} = \frac{t^{6n-3}+t^{3k-6n+3}}{(1-t^6)(1-t^{3k})}
\end{equation}

In summary, the constrained sum of $t^{3n_1+3n_2}$ for fields having $j_2=\pm 1/2$ is

\begin{align}
\label{eq:appendixzkf1}
F'^n_{\mathbb{Z}_k,j_2=\pm 1/2}&=
\frac{6(t^3+t^{3k-3})}{(1-t^6)(1-t^{3k})} \quad n=0, k/2\\
\label{eq:appendixzkf2}
F'^n_{\mathbb{Z}_k,j_2=\pm 1/2}&=
\frac{2(t^3+t^{3k-3}+t^{6n+3}+t^{3k-6n-3}+t^{6n-3}+t^{3k-6n+3})}{(1-t^6)(1-t^{3k})} \quad n\neq 0, k/2
\end{align}

\section{$\text{Dic}_k$ Index Calculation}
\label{sec:appendix2}
In this appendix we work out the $\text{Dic}_k$ single letter index in detail. First, we deal with the abelian holonomies. As discussed in section~\ref{sec:d}, there are four $SU(2)$ abelian holonomies where $g_j(r)$ and $g_j(s)$ ($j=1,2,3,4$) take values in $\pm 1$. Since all of the fields transform in the ajdoint of $SU(2)$, the four abelian holonomies act as identity on the fields. We first compute the contribution to the index by the $j_2=0$ fields. A general single letter operator can be written as
\begin{equation}
    \Psi = \sum_{n_1n_2l}C_{n_1n_2l}\partial_{++}^{n_1} \partial_{+-}^{n_2} \Phi_l
\end{equation}
where $n_1,n_2 \geq 0$ and $l=-1,0,1$ is the $SU(2)$ adjoint index. The operator has to satisfy two constraints~\eqref{eq:constraint} given by the generators $r$ and $s$. The two constraints are
\begin{align}
    \sum_{n_1n_2l} \exp((n_1-n_2)\pi i/k)C_{n_1n_2l}\partial_{++}^{n_1} \partial_{+-}^{n_2} \Phi_l &=\sum_{n_1n_2l}C_{n_1n_2l}\partial_{++}^{n_1} \partial_{+-}^{n_2} \Phi_l \\
    \sum_{n_1n_2l} \exp(n_2\pi i)C_{n_1n_2l}\partial_{++}^{n_2} \partial_{+-}^{n_1} \Phi_l &=\sum_{n_1n_2l}C_{n_1n_2l}\partial_{++}^{n_1} \partial_{+-}^{n_2} \Phi_l
\end{align}

The first equation implies that for $C_{n_1n_2l}$ to be nonvanishing, we need
\begin{equation}
\label{eq:appendixdic1}
    n_1-n_2 = 0\mod{2k} 
\end{equation}

Exchanging the label $n_1$ with $n_2$, we see that the second equation implies
\begin{equation}
\label{eq:appendixdic2}
    \exp(n_1\pi i) C_{n_2n_1 l} \exp(n_1\pi i)  = C_{n_1n_2 l} 
\end{equation}

Equation~\eqref{eq:appendixdic2} shows that we only need to sum over $n_1 \geq n_2$ subject to the condition~\eqref{eq:appendixdic1}, since $C_{n_1n_2l}$ is a function of $C_{n_2n_1l}$. For $n_1=n_2$, equation~\eqref{eq:appendixdic2} shows that we need to sum over $n_1=n_2 = 2\mathbb{N}$ only, since an odd value of $n_1$ would lead to a vanishing $C_{n_1n_2 l}$. Therefore, the constrained sum on $t^{3n_1+3n_2}$ becomes
\begin{align}
\label{eq:appendixfdickabelianj0}
    F^{\text{abelian}}_{\text{Dic}_k, j_2=0}\equiv 3\sum t^{3n_1+3n_2} = \frac{3t^{6k}}{(1-t^6)(1-t^{6k})} + \frac{3}{1-t^{12}}
\end{align}
where the prefactor comes from the fact that $l$ can take three values. 

Now we discuss the case for $j_2=\pm 1/2$ fields. The most general operator for such fields $\Phi_{\mu l}$ is
\begin{equation}
    \Psi' = \sum C_{n_1n_2\mu l}  \partial_{++}^{n_1} \partial_{+-}^{n_2} \Phi_{\mu l}
\end{equation}
where $\mu=\pm 1$ is the $SU(2)$ doublet index since the fields, having $j_2=\pm 1/2$, now form a doublet. The difference between this case and the previous $j_2=0$ case is that now, the geometric action of $r$ and $s$ will be affected by the $\mu$ index. The $r$ and $s$ constraints on $\Psi'$ are
\begin{align}
    \sum \exp((n_1-n_2+\mu)\pi i/k) C_{n_1n_2\mu l}  \partial_{++}^{n_1} \partial_{+-}^{n_2} \Phi_{\mu l}&=\sum C_{n_1n_2\mu l}  \partial_{++}^{n_1} \partial_{+-}^{n_2} \Phi_{\mu l} \\
    \sum \exp(n_2 \pi i) (-1)^{(\mu-1)/2}C_{n_1n_2\mu l}  \partial_{++}^{n_2} \partial_{+-}^{n_1} \Phi_{-\mu l}&=\sum C_{n_1n_2\mu l}  \partial_{++}^{n_1} \partial_{+-}^{n_2} \Phi_{\mu l}
\end{align}

They imply
\begin{align}
    \label{eq:appendixdic3}
    n_1-n_2+\mu &= 0\mod{2k} \\
    \label{eq:appendixdic4}
    \exp(n_1 \pi i) (-1)^{-(\mu+1)/2} C_{n_2n_1,-\mu,l} &= C_{n_1n_2 \mu l}
\end{align}

Equation~\eqref{eq:appendixdic4} suggests that we need to sum over all $n_1, n_2$ pairs such that equation~\eqref{eq:appendixdic3} holds for $\mu=1$, since the coefficient for $\mu=-1$ is related to that for $\mu=1$ via equation~\eqref{eq:appendixdic4}. Therefore, the constrained sum on $t^{3n_1+3n_2}$ becomes
\begin{equation}
\label{eq:appendixfdickabelianj1}
    F^{\text{abelian}}_{\text{Dic}_k, j_2=\pm 1/2} \equiv 3\sum t^{3n_1+3n_2} = \frac{3t^3+3t^{6k-3}}{(1-t^6)(1-t^{6k})}
\end{equation}

At this point, we have computed the contribution of $\partial_{+\pm}$ to the index. We also need to add the contribution to the index by the base fields (i.e. those acted on by the derivative operators). Looking up Table~\ref{table:n=4bpstable}, we can read off the $t^{2(D+j_1)}$ value for each field, and the total single letter index is 
\begin{equation}
    \hat{I}^{\text{abelian}}_{\text{Dic}_k} =F^{\text{abelian}}_{\text{Dic}_k, j_2=0}(3t^2-3t^4+2t^6)-F^{\text{abelian}}_{\text{Dic}_k, j_2=\pm 1/2}t^3
\end{equation}
and its small $\beta$ expansion up to first order is
\begin{equation}
    \hat{I}^{\text{abelian}}_{\text{Dic}_k}= 3-\frac{\beta}{3k} + O(\beta^2)
\end{equation}

Next, we work out the single letter index for the nonabelian holonomies $g'_n$, $n=1,3,5,...,k-1$, as discussed in section~\ref{sec:d}. In this case, the holonomies will act nontrivially on the fields. Recall from section~\ref{sec:d} that 
\begin{align*}
g_n'(r)&=\begin{pmatrix}
\exp(n\pi i/k) & 0 \\
0 & \exp(-n\pi i/k) 
\end{pmatrix} \quad n=1,3,5,...,k-1 \\
g'_n(s)&=
\begin{pmatrix}
0 & 1 \\
-1 & 0
\end{pmatrix} \quad n=1,3,5,...,k-1
\end{align*}

The adjoint action of the $n$th holonomy on the fields is
\begin{align}
    \label{eq:appendixdic5}
    g'_n(r)(\Phi_0, \Phi_1, \Phi_{-1})g'^{-1}_n(r) &= (\Phi_0, e^{2n\pi i/k}\Phi_1, e^{-2n\pi i/k} \Phi_{-1}) \\
    \label{eq:appendixdic6}
    g'_n(s)(\Phi_0, \Phi_1, \Phi_{-1})g'^{-1}_n(s) &= -(\Phi_0, \Phi_{-1}, \Phi_1)
\end{align}

In particular, $g'_n(s)$ exchanges the 1 and the $-1$ components of the field. 

As we did before, we first discuss the contribution to the index by fields having $j_2=0$. The most general single letter operator one can form is as before
\begin{equation}
    \Psi = \sum_{n_1n_2l}C_{n_1n_2l}\partial_{++}^{n_1} \partial_{+-}^{n_2} \Phi_l
\end{equation}

Using equations~\eqref{eq:appendixdic5} and \eqref{eq:appendixdic6} in equation~\eqref{eq:constraint}, we find the constraints \begin{align}
    \sum_{n_1n_2l} \exp((n_1-n_2)\pi i/k)C_{n_1n_2l}\partial_{++}^{n_1} \partial_{+-}^{n_2} \Phi_l &=\sum_{n_1n_2l} \exp(2\pi nli/k)C_{n_1n_2l}\partial_{++}^{n_1} \partial_{+-}^{n_2} \Phi_l \\
    \sum_{n_1n_2l} \exp(n_2\pi i)C_{n_1n_2l}\partial_{++}^{n_2} \partial_{+-}^{n_1} \Phi_l &=-\sum_{n_1n_2l}C_{n_1n_2l}\partial_{++}^{n_1} \partial_{+-}^{n_2} \Phi_{-l}
\end{align}

They imply 
\begin{align}
    \label{eq:appendixdic7}
    n_1-n_2 - 2nl &= 0\mod{2k} \\
    \label{eq:appendixdic8}
    -e^{n_1 \pi i}C_{n_2n_1,-l} &= C_{n_1n_2l}
\end{align}

For the case $l=0$, equation~\eqref{eq:appendixdic8} tells us to sum over all $n_1>n_2$ pairs satisfying equation~\eqref{eq:appendixdic7}, and for $n_1=n_2$ we need to sum over odd $n_1$. The constrained $t^{3n_1}t^{3n_2}$ sum is thus
\begin{equation}
    \sum t^{3n_1}t^{3n_2} = \frac{t^{6k}}{(1-t^6)(1-t^{6k})}+ \frac{t^6}{1-t^{12}} 
\end{equation}

For the case $l=1$, we need to sum over all $(n_1,n_2)$ pairs that satisfy equation~\eqref{eq:appendixdic7}, and according to equation~\eqref{eq:appendixdic8} this will take care of the $l=-1$ case automatically. The constrained $t^{3n_1}t^{3n_2}$ sum in this case is
\begin{equation}
    \sum t^{3n_1}t^{3n_2} = \frac{t^{6n}+t^{6k-6n}}{(1-t^6)(1-t^{6k})} 
\end{equation}

Overall, the constrained sum is
\begin{equation}
\label{eq:appendixdicf0}
    F^{n}_{\text{Dic}_k, j_2=0} = \frac{t^{6k}+t^{6n}+t^{6k-6n}}{(1-t^6)(1-t^{6k})}+ \frac{t^6}{1-t^{12}}
\end{equation}
 
Now consider the fields that have $j_2=\pm 1/2$. The most general operator built out of such fields is 
\begin{equation}
    \Psi' = \sum C_{n_1n_2\mu l}  \partial_{++}^{n_1} \partial_{+-}^{n_2} \Phi_{\mu l}
\end{equation}
where now we have a doublet index $\mu=\pm 1$. The constraints are
\begin{align}
    \sum \exp((n_1-n_2+\mu)\pi i/k) C_{n_1n_2\mu l}  \partial_{++}^{n_1} \partial_{+-}^{n_2} \Phi_{\mu l}&=\sum C_{n_1n_2\mu l} \exp(2\pi nli/k) \partial_{++}^{n_1} \partial_{+-}^{n_2} \Phi_{\mu l} \\
    \sum \exp(n_2 \pi i) (-1)^{(\mu-1)/2} C_{n_1n_2\mu l}  \partial_{++}^{n_2} \partial_{+-}^{n_1} \Phi_{-\mu l}&=-\sum C_{n_1n_2\mu l}  \partial_{++}^{n_1} \partial_{+-}^{n_2} \Phi_{\mu,-l}
\end{align}

They imply
\begin{align}
    \label{eq:appendixdic9}
    n_1-n_2+\mu-2nl &= 0\mod{2k} \\
    \label{eq:appendixdic10}
    -e^{n_1\pi i} (-1)^{-(\mu+1)/2} C_{n_2n_1,-\mu,-l} &= C_{n_1n_2\mu l}
\end{align}

Equation~\eqref{eq:appendixdic10} suggests that we only need to sum over all $(n_1,n_2)$ pairs satisfying equation~\eqref{eq:appendixdic9} for each of the three independent cases: $(\mu=-1,l=0)$, $(\mu=-1, l=1)$, $(\mu=-1, l=-1)$. Adding up the contributions to $t^{3n_1+3n_2}$ from all three cases, one obtains
\begin{equation}
\label{eq:appendixdicf1}
    F^{n}_{\text{Dic}_k, j_2=\pm 1/2}=\frac{t^3+t^{6k-3}+t^{6n+3}+t^{6k-6n-3}+t^{6n-3}+t^{6k-6n+3}}{(1-t^6)(1-t^{6k})}
\end{equation}

Therefore, the total single letter index for the $n$th ground state Wilson line is
\begin{equation}
    \hat{I}^{n}_{\text{Dic}_k} =F^{n}_{\text{Dic}_k, j_2=0}(3t^2-3t^4+2t^6)-F^{n}_{\text{Dic}_k, j_2=\pm 1/2}t^3
\end{equation}

Expanding in $\beta$ up to first order, we get
\begin{equation}
    \hat{I}^{n}_{\text{Dic}_k} =-\frac{\beta}{3k} +O(\beta^2)
\end{equation}

\section{Ground State t' Hooft Lines on $S^3/\mathbb{Z}_k$}
\label{sec:appendixmagneticzk}

As mentioned in section~\ref{sec:sdualityzk}, for $k$ odd the t' Hooft line solutions are the same as the Wilson line solutions. For $k$ even, the t' Hooft line solutions are
\begin{equation}
    \tilde{g}_n(r) =\begin{pmatrix}
        e^{\pi in /k} & 0 \\
        0 & e^{-\pi in /k}
    \end{pmatrix}, \quad n=0,1,...,k/2
\end{equation}

Using the method from the previous appendices, we compute the $T$ functions that go into the computation of single letter indices as in equation~\eqref{eq:thooftindexzk}. It turns out that the $n=0$ case is different from the rest, so we list the $T$ functions separately for both cases.

\begin{align}
    \label{eq:appendixzkmag1}
    T^0_{\mathbb{Z}_k,j_2=0} &= 3\frac{1+t^{3k}}{(1-t^6)(1-t^{3k})} \\
    \label{eq:appendixzkmag2}
    T^n_{\mathbb{Z}_k,j_2=0} &= \frac{1+t^{3k}+2(t^{3n}+t^{3k-3n})}{(1-t^6)(1-t^{3k})}, \quad n\neq 0
\end{align}

\begin{align}
    \label{eq:appendixzkmag3}
    T^0_{\mathbb{Z}_k,j_2=\pm 1/2} &= 6\frac{t^3+t^{3k-3}}{(1-t^6)(1-t^{3k})}\\
    \label{eq:appendixzkmag4}
    T^n_{\mathbb{Z}_k,j_2=\pm 1/2} &= 2\frac{t^3+t^{3k-3}+t^{3n-3}+t^{3k+3-3n}+t^{3n+3}+t^{3k-3-3n}}{(1-t^6)(1-t^{3k})}, \quad n\neq 0
\end{align}

\section{Ground State t' Hooft Lines on $S^3/\text{Dic}_k$}
\label{sec:appendixmagneticdick}

In this appendix, we first show that the number of $SU(2)/\mathbb{Z}_2$ t' Hooft line ground states is the same as that of the $SU(2)$ Wilson line ground states on $S^3/\text{Dic}_k$, and then tabulate the $T$ functions used in the computation of single letter index. Let $\tilde{g}(r)$ and $\tilde(g)(s)$ denote a t' Hooft line representation for the generators $r$ and $s$. As mentioned in section~\ref{sec:sduality} they must satisfy
\begin{equation}
\label{eq:appendixmagneticdickconstraint}
    \tilde{g}(r)^{2k} = \pm \textbf{1}, \quad \tilde{g}(r)^k = \pm \tilde{g}(s)^2, \quad \tilde{g}^{-1}(s)\tilde{g}(r)\tilde{g}(s)= \pm \tilde{g}^{-1}(r)
\end{equation}

We choose to diagonalize $\tilde{g}(r)$. It is easy to see that 
\begin{equation}
    \tilde{g}_n(r) = 
    \begin{pmatrix}
        e^{\pi in/2k} & 0\\
        0 & e^{-\pi in/2k}
    \end{pmatrix}, \quad n=0,...,k
\end{equation}
satisfies the first equation in~\eqref{eq:appendixmagneticdickconstraint}. We do not need $n > k$, since, in the same spirit as the $\mathbb{Z}_k$ case, these solutions are identified with the $n\leq k$ ones under multiplication by $-1$ and the Weyl group exchanging the diagonal elements in the matrix. Let us parameterize $\tilde{g}(s)$ by $\tilde{g}(s)= \exp(i\theta \hat{n}\cdot\vec{\sigma})$ where $\vec{\sigma}$ is a vector of the three Pauli matrices. The second and the third equation in~\eqref{eq:appendixmagneticdickconstraint} implies
\begin{align}
    \cos(\pi n/2) \textbf{1} + i \sin(\pi n/2) \sigma_3 &= \pm (\cos (2\theta) \textbf{1} + i \sin(2\theta) \hat{n} \cdot \vec{\sigma}) \\
    \cos(\pi n/2k) \textbf{1} +i\sin(\pi n/2k) \tilde{g}(s)^{-1} \sigma_3 \tilde{g}(s)&=  \pm \cos(\pi n/2k) \textbf{1} -i\sin(\pi n/2k)\sigma_3 
\end{align}

There are three cases to consider. 
\begin{itemize}
    \item Case $n=0$. In this case there are two solutions to $\tilde{g}(s)$:
    \begin{equation}
        \tilde{g}(s)=\begin{pmatrix}
            1 & 0 \\
            0 & 1
        \end{pmatrix}
        ,
        \begin{pmatrix}
            i & 0 \\
            0 & -i
        \end{pmatrix}
    \end{equation}
    It might seem that there can be another solution for $\tilde{g}(s)$, namely
    \begin{equation}
        \tilde{g}(s)=\begin{pmatrix}
            0 & 1 \\
            -1 & 0
        \end{pmatrix}
    \end{equation}
    But this $\tilde{g}(r), \tilde{g}(s)$ pair can be obtained from the previous one by an $SU(2)$ conjugation.

    \item Case $n=k$. Recall that $k$ is assumed even, so in this case there are three solutions to $\tilde{g}(s)$:
    \begin{equation}
        \tilde{g}(s)=\begin{pmatrix}
            1 & 0 \\
            0 & 1
        \end{pmatrix}
        ,
        \begin{pmatrix}
            i & 0 \\
            0 & -i
        \end{pmatrix}
        \begin{pmatrix}
            0 & 1 \\
            -1 & 0
        \end{pmatrix}
    \end{equation}
    Although the last of the above solutions by itself can be obtained from conjugating the previous solution, the conjugation would change $\tilde{g}(r)$ in this case, making the last solution unique. 

    \item Case $0<n<k$. In this case it is easy to see that there is no solution for odd $n$. The only solutions are

    \begin{align}
    \tilde{g}_n(r) &= 
    \begin{pmatrix}
        e^{\pi in/2k} & 0\\
        0 & e^{-\pi in/2k}
    \end{pmatrix}, \quad n=2,4,...,k-2 \\
    \tilde{g}_s(r) &= 
    \begin{pmatrix}
        0 & 1\\
        -1 & 0
    \end{pmatrix}, \quad n=2,4,...,k-2
    \end{align}
\end{itemize}

In total, there are $4+k/2$ solutions, in agreement with the number of Wilson line ground states. Note that the five solutions from the first two cases are universal. They are valid for all $k$. 

Next, we tabulate the $T$ functions used in equation~\eqref{eq:thooftdickindex}, first for the five universal solutions and then for the $k/2-1$ solutions dependent on $k$. We denote the $T$ functions as $T^j, j=1,...,5$ for the five universal solutions and as $T'^n, n=2,4,...,k-2$ for the $k/2-1$ solutions.

\begin{align}
    T^1_{\text{Dic}_k, j_2=0} &= \frac{3t^{6k}}{(1-t^6)(1-t^{6k})}+ \frac{3}{1-t^{12}} \\
    T^1_{\text{Dic}_k, j_2=\pm 1/2} &= \frac{3t^{3}+3t^{6k-3}}{(1-t^6)(1-t^{6k})} \\
    T^2_{\text{Dic}_k, j_2=0} &= \frac{3t^{6k}}{(1-t^6)(1-t^{6k})}+ \frac{1+2t^6}{1-t^{12}} \\
    T^2_{\text{Dic}_k, j_2=\pm 1/2} &= \frac{3t^{3}+3t^{6k-3}}{(1-t^6)(1-t^{6k})} \\
    T^3_{\text{Dic}_k, j_2=0} &= \frac{t^{6k}+2t^{3k}}{(1-t^6)(1-t^{6k})}+ \frac{1}{1-t^{12}} \\
    T^3_{\text{Dic}_k, j_2=\pm 1/2} &= \frac{t^{3}+t^{6k-3}+2(t^{3k-3}+t^{3k+3})}{(1-t^6)(1-t^{6k})} \\
    T^4_{\text{Dic}_k, j_2=0} &= \frac{t^{6k}+2t^{3k}}{(1-t^6)(1-t^{6k})}+ \frac{1}{1-t^{12}} \\
    T^4_{\text{Dic}_k, j_2=\pm 1/2} &= \frac{t^{3}+t^{6k-3}+2(t^{3k-3}+t^{3k+3})}{(1-t^6)(1-t^{6k})} \\
    T^5_{\text{Dic}_k, j_2=0} &= \frac{t^{6k}+2t^{3k}}{(1-t^6)(1-t^{6k})}+ \frac{t^6}{1-t^{12}} \\
    T^5_{\text{Dic}_k, j_2=\pm 1/2} &= \frac{t^{3}+t^{6k-3}+2(t^{3k-3}+t^{3k+3})}{(1-t^6)(1-t^{6k})} 
\end{align}

\begin{align}
    T'^n_{\text{Dic}_k, j_2=0} &= \frac{t^{6k}+t^{3n}+t^{6k-3n}}{(1-t^6)(1-t^{6k})}+ \frac{t^6}{1-t^{12}}, \quad n=2,4,...,k-2\\
    T'^n_{\text{Dic}_k, j_2=\pm 1/2} &= \frac{t^3+t^{6k-3}+t^{3n-3}+t^{6k+3-3n}+t^{3n+3}+t^{6k-3-3n}}{(1-t^6)(1-t^{6k})}, \quad n=2,4,...,k-2
\end{align}

\end{document}